\documentclass[%
preprint,
notitlepage,
]{revtex4-2}

\usepackage{amsmath,amssymb}
\usepackage[colorlinks=true, allcolors=blue]{hyperref}
\usepackage{graphicx}
\usepackage{dcolumn}
\usepackage{bm}

\begin{document}


\title{Transport statistics in asymmetric coupled chaotic dots in series: Brownian motion model}

\author{Ashutosh Dheer}
\email{ad326@snu.edu.in}

\author{Santosh Kumar}%
\thanks{Deceased, 18 October 2024.}
\affiliation{%
 Department of Physics, Shiv Nadar Institution of Eminence, Gautam Buddha Nagar, Greater Noida, India
}%

\author{Jayendra N. Bandyopadhyay}
\affiliation{
 Department of Physics, Birla Institute of Technology and Science Pilani, Pilani Campus, Vidya Vihar, Pilani, Rajasthan 333031, India
}%

\begin{abstract}
We study statistics of transport properties in a system of two quantum chaotic cavities coupled in series through a tunable junction and connected to reservoirs with asymmetric leads. Within the random matrix theory (RMT) framework and in the absence of time-reversal symmetry, the scattering matrix $S$ undergoes a crossover from two independent Circular Unitary Ensembles (2CUE) to a single CUE as the inter-cavity coupling increases. Modeling this crossover using Dyson’s Brownian motion approach, with the junction width acting as a time-like parameter, we derive closed-form expressions for the time-dependent averages and variances of conductance and shot noise power for arbitrary lead asymmetry. These analytical results reduce to known expressions in the symmetric and fully coupled limits. The mapping between the Brownian time parameter and the physical junction width is found to be observable-dependent, reflecting the different eigenfunction correlations probed by conductance and shot noise power, and this dependence is examined and separately calibrated for each observable. Validating against random-matrix simulations and tight-binding transport calculations using Kwant, we find quantitative agreement for ensemble-averaged observables throughout the crossover, while variances, being more sensitive to microscopic fluctuations, show deviations from the Brownian-motion predictions.
\end{abstract}

\maketitle
\newpage


\section{\label{sec:level1}Introduction}

Quantum transport in mesoscopic systems provides a natural setting for studying universal statistical properties of complex quantum systems. In ballistic chaotic cavities, transport properties such as conductance and shot noise power exhibit strong sample-to-sample fluctuations, yet their statistical features are largely independent of microscopic details. Random matrix theory (RMT)~\cite{mehta2004random, forrester2010log, Guhr1998} offers a natural framework to capture this universality, with the scattering matrix $S$ modeled by Dyson’s circular ensembles according to the symmetries of the system. Early theoretical foundations of universal transport fluctuations in chaotic cavities were established using semiclassical and random-matrix approaches~\cite{Lee1985, Jalabert1994, Baranger1994, Beenakker1997, Alhassid2000}, with subsequent studies extending these frameworks to more general transport settings~\cite{Savin2006, Sommers2007, Khoruzhenko2009, Poli2009, Vidal2012, RodrguezPrez2013, Dheer2022}.

\begin{figure}[!htb]
    \centering
    \includegraphics[width=9 cm]{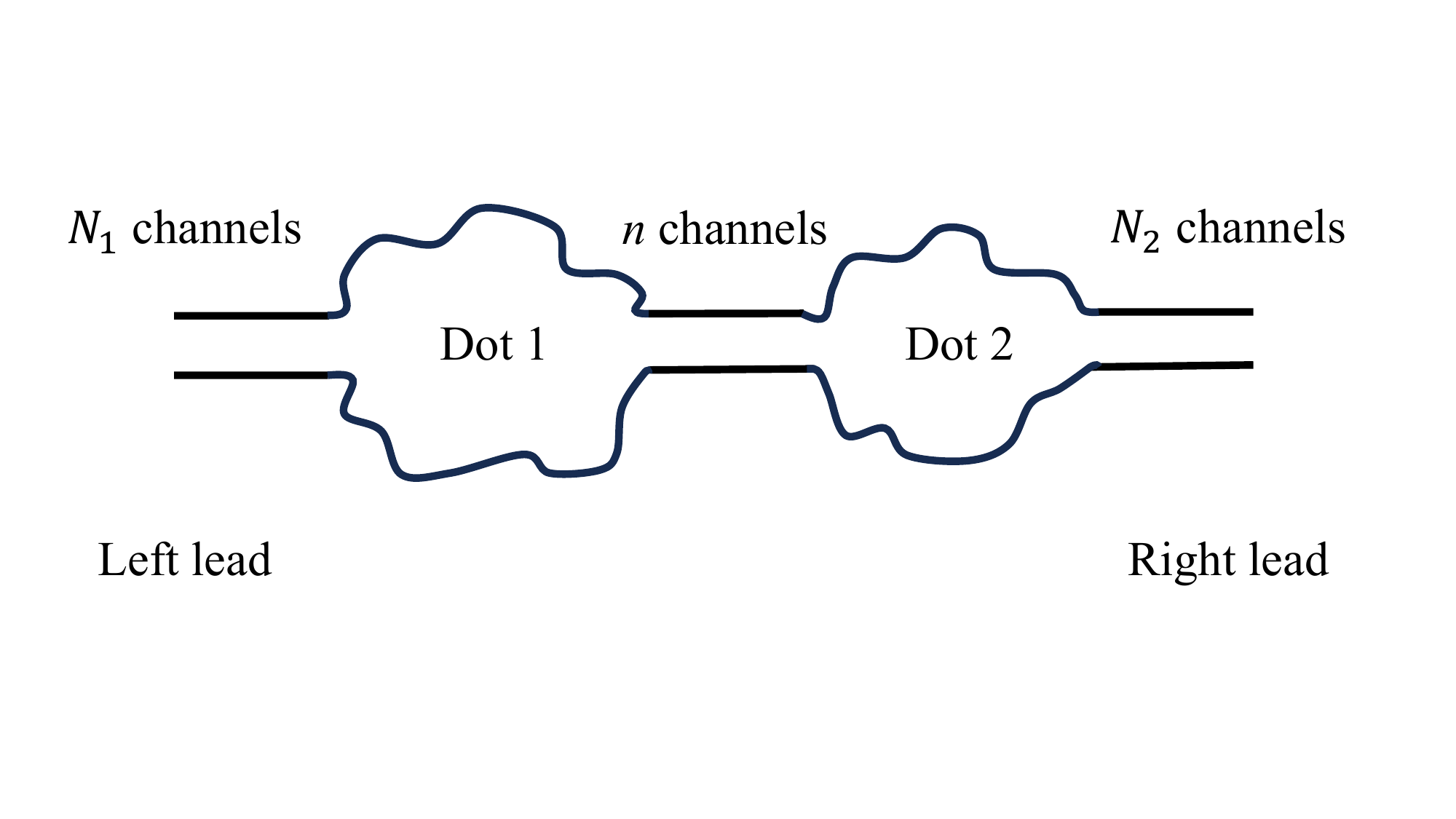}
    \caption{Schematic of two chaotic cavities coupled via a junction with $n$ channels and connected to reservoirs through leads supporting $N_{1}$ and $N_{2}$ channels. The symmetric case $N_{1}=N_{2}$ was studied in Ref.~\cite{Frahm1995brownian}, here we treat the general asymmetric configuration.}
    \label{fig:Coupled cavity}
\end{figure}

For a two-terminal chaotic cavity connected to reservoirs by ideal leads supporting $N_1$ and $N_2$ channels, the transmission eigenvalues follow a Jacobi ensemble joint distribution. The transmission eigenvalues $\{T_{i}\}$ are the eigenvalues of $tt^{\dagger}$, where $t$ is the transmission block of $S$. Coupling multiple chaotic cavities can lead to crossover phenomena in the symmetry of the effective scattering matrix. A notable example is a system of two chaotic cavities coupled in series through a tunable junction (see Fig.~\ref{fig:Coupled cavity}) and subjected to a magnetic field, such that time-reversal symmetry is broken. In this case, the scattering matrix undergoes a crossover from two independent Circular Unitary Ensembles (2CUE) to a single CUE as the coupling between the cavities increases. (We note that this terminology refers to the modeling of the $S$-matrix statistics by RMT ensembles, which is well established to accurately capture short-range spectral fluctuations of quantum chaotic systems; deviations from RMT universality at larger scales are known to arise from non-universal semiclassical contributions of short periodic orbits~\cite{Berry1985}.) Frahm and Pichard studied this crossover~\cite{Frahm1995brownian} using a Brownian motion model, computing the average and variance of conductance for the symmetric lead configuration $N_1 = N_2$.

However, transport statistics for this crossover remain unexplored when the two cavities are connected to reservoirs with leads having unequal numbers of channels, $N_1 \neq N_2$. From an experimental perspective, continuously tunable coupling between quantum dots can be realized using gate-defined constrictions, quantum point contacts, and lithographically controlled junctions. Such setups naturally generate crossover regimes between effectively disconnected cavities and fully merged composite systems. Understanding how transport observables evolve across this crossover is therefore important not only for testing random-matrix predictions, but also for interpreting experiments on coupled mesoscopic structures. In particular, asymmetric lead configurations represent generic experimental situations and can significantly modify transport statistics compared to the symmetric case. In this work, we address this gap by obtaining transport statistics using Dyson’s Brownian motion method for the Jacobi ensemble, with the inter-cavity coupling acting as a time-like parameter. We derive closed-form expressions for the time-dependent averages and variances of conductance and shot noise power across the full 2CUE $\rightarrow$ 1CUE crossover. Our results reduce to known expressions in the symmetric and fully coupled limits.

We validate our analytical predictions using both random-matrix simulations and tight-binding transport calculations performed with the Kwant package. We find excellent agreement for ensemble-averaged observables, confirming that the Brownian motion framework captures the essential features of the crossover. At the level of fluctuations, however, clear deviations emerge, in particular, the variances show enhanced sensitivity to microscopic details. We clarify the limitations of the Brownian motion description for variances and higher-order statistics. In addition, we highlight that the connection between the Brownian time parameter $\tau$ and the physical junction width $n$ is established through a phenomenological argument, and this mapping is found in practice to be dependent on the observable used to calibrate it.

The paper is organized as follows. In Sec.~\ref{sec:level2} we introduce the Brownian motion framework for the asymmetric crossover. Sec.~\ref{sec:level3} presents analytical results for time-dependent transport statistics. In Sec.~\ref{sec:level4} we compare these predictions with numerical simulations and discuss the mapping between junction width and Brownian time. Finally, Sec.~\ref{sec:level5} summarizes our findings and their implications for mesoscopic transport.

\section{\label{sec:level2}Brownian motion ensemble of quantum transport}

The joint probability density function (JPDF) of $N$ eigenvalues $\{x_{1},\dots,x_{N}\}$ of the classical invariant ensembles of RMT has the form~\cite{mehta2004random, forrester2010log}
\begin{align}
    P^{\beta}(\{x\})=\frac{1}{\mathcal{Z}_{N}^{\beta}}~|\Delta_{N}(\{x\})|^{\beta}\prod_{k=1}^{N}w(x_{k})\label{jpdf}.
\end{align}
$(\mathcal{Z}_{N}^{\beta})^{-1}$ is the normalization constant which can be evaluated using, for example, the well-known Selberg integral. The factor $\Delta_{N}(\{x\})=\prod_{j>k}^{N}(x_{j}-x_{k})$ is the Vandermonde determinant, $w(x)$ is the weight function and the Dyson index, $\beta\in\{1,2,4\}$ corresponds to the orthogonal, unitary, and symplectic symmetry classes, respectively. The choice of Jacobi weight
\begin{align}
    w_{J}^{(\beta)}(x)=(1-x)^{\beta(a+1)-1}\,(1+x)^{\beta(b+1)-1}\label{weight}
\end{align}
with
\begin{align}
    a,b>-1 \text{~~and~~} x\in[-1,1]
\end{align}
defines the Jacobi ensemble, $P_{J}^{\beta}(\{x\})$. It appears naturally in quantum transport problems~\cite{Beenakker1997, Damle2011, Forrester2006}, wireless communication~\cite{Dar2013, Laha2021}, multivariate statistics~\cite{Constantine1963, Collins2005, Johnstone2008}, and in integrable systems, most notably through its connection with the Painlev{\'e} VI equation~\cite{Haine1999}.

In the Landauer–B\"{u}ttiker framework~\cite{Landauer1996, Bttiker1990, Fisher1981}, transport through a phase-coherent conductor is fully characterized by the transmission eigenvalues $\{T_{i}\}$, which are the eigenvalues of the Hermitian matrix $tt^{\dagger}$, with $t$ the transmission block of the scattering matrix $S$. These eigenvalues lie in the interval $[0,1]$ and represent the transmission probabilities of independent transport channels. For a two-terminal chaotic mesoscopic cavity connected to reservoirs with ideal leads supporting $\{N_{1}, N_{2}\}$ channels, there are $N~(=\min(N_{1},N_{2}))$ transmission eigenvalues $\{T_{1},\dots,T_{N}\}$, whose JPDF takes the form~\cite{Beenakker1997}
\begin{align}
    P^{\beta}(\{T\})\propto\prod_{i<j}^{N}|T_{i}-T_{j}|^{\beta}\prod_{k=1}^{N}T_{k}^{\frac{\beta}{2}(N_{\Delta}+1-2/\beta)}\label{2terminaljpdf}
\end{align}
with $N_\Delta=|N_{1}-N_{2}|$. This distribution can be obtained from the JPDF of the Jacobi ensemble, $P_{J}^{\beta}(\{x\})$, via the transformation
\begin{align}\label{transformations}
    x=1-2T,\qquad a=\frac{N_{\Delta}-1}{2},\qquad b=\frac{1}{\beta}-1.
\end{align}
In what follows, we work in the $x$-variable Jacobi representation~\eqref{weight} for convenience, and perform the inverse substitution $T_{i}=(1-x_{i})/2$ only at the final stage to express transport statistics in terms of transmission eigenvalues.

The JPDF in Eq.~\eqref{jpdf} admits a static Coulomb gas interpretation~\cite{Dyson1962statistical}, in which eigenvalues behave as logarithmically repelling charged particles confined by an external potential. Dyson showed that this equilibrium picture can be extended dynamically, with the eigenvalues evolving stochastically in a fictitious time $\tau$ according to Brownian motion~\cite{Dyson1962}. As $\tau$ increases, the probability distribution of eigenvalues $P(\{x\};\tau)$ diffuses in the eigenvalue configuration space, constrained by the symmetries of the ensemble, and eventually relaxes to the equilibrium distribution $P^{\beta}(\{x\})$ as $\tau\rightarrow\infty$. The time-dependent JPDF $P(\{x\};\tau)$ then satisfies a Fokker–Planck equation
\begin{align}
    \frac{\partial P(\{x\};\tau)}{\partial\tau}=-\mathcal{L}\,P(\{x\};\tau)\label{fokker-planck}
\end{align}
where the Fokker-Planck operator takes the form
\begin{align}
    \mathcal{L}=\sum_{i=1}^{N}\frac{\partial}{\partial x_{i}}\,e^{-\beta W}\frac{\partial}{\partial x_{i}}f(x_{i})\,e^{\beta W},
\end{align}
$W$ is the effective potential
\begin{align}
W=-\sum_{i<j}^{N}\ln|x_i-x_j|-\frac{1}{\beta}\sum_{k=1}^N\ln w(x_k)
\end{align}
and $f(x)$ is related to the local mobility or inverse friction coefficient. For the Jacobi ensemble, it takes the form, $f(x)=-(1-x)^{2}$. The equilibrium condition $\mathcal{L}P=0$ ensures that the stationary solution is indeed the JPDF $P^{\beta}(\{x\})$. Brownian-motion and Fokker--Planck descriptions have also found important applications in quantum-transport. In particular, the Dorokhov--Mello--Pereyra--Kumar (DMPK) equation describes the evolution of the joint distribution of transmission eigenvalues with increasing length in quasi-one-dimensional disordered multichannel wires~\cite{Dorokhov1982, Mello1988, Shukla2005}. Brownian-motion approaches have also been used to study parametric correlations of spectral and transport quantities as system parameters are varied~\cite{Pandey1981, Mehta1983spacing, Pandey1983, French1988, Pandey1991, Pandey1993, Beenakker1993Nonlogarithmic, Beenakker1993Brownian, Pandey1995, Frahm1995magneto}.

The Fokker-Planck equation~\eqref{fokker-planck} can be mapped exactly onto a Schr\"{o}dinger-type equation in imaginary time $i\tau$ with the associated Hamiltonian taking the form of the Calogero-Sutherland Hamiltonian~\cite{Pandey1995}. For $\beta=2$, this Hamiltonian describes $N$ noninteracting fermions; specializing to the Jacobi ensemble, the corresponding single-particle eigenfunctions are given by weighted Jacobi polynomials
\begin{align}
    [w^{(2)}(x)]^{1/2}\,P_{\mu}^{(2a+1,0)}(x),
\end{align}
(the second Jacobi parameter vanishes for $\beta=2$) and eigenvalues
\begin{align}
    \varepsilon_{\mu}=\mu(\mu+2a+3),
\end{align}
while the many-body ground state energy is $\mathcal{E}_{0}=N(N-1)(N+3a+4)/6$. The Fokker-Planck equation can be formally inverted using a Green's function approach, as outlined in~\cite{Kumar2011}, yielding
\begin{align}
    P(\{x\};\tau)=\int_{\mathbf{R}}\dots\int_{\mathbf{R}}\mathcal{P}(\{x\};\tau|\{y\};0)\,P(\{y\};0)\,dy_{1}\dots dy_{N}\label{Integraleqn}
\end{align}
where $\mathcal{P}(\{x\};\tau|\{y\};0)$ serves as the Green’s function (or the transition kernel), representing the conditional joint probability density for the system to evolve from an initial eigenvalue configuration $\{y\}$ at $\tau = 0$ to a final configuration $\{x\}$ at time $\tau$, given by~\cite{Pandey1991, Pandey2004}
\begin{align}
    P(\{x\};\tau|\{y\};0)=\frac{e^{\mathcal{E}_{0}\tau}}{N!}\frac{\Delta_{N}(\{x\})}{\Delta_{N}(\{y\})}\prod_{i=1}^{N}(1-x_{i})^{2a+1}\,\text{det}[K_{N}(x_{j},y_{k};\tau)]_{j,k=1,\dots,N}\label{Green}
\end{align}
with the single-particle kernel
\begin{align}
    K_{N}(x,y;\tau)=\sum_{\mu=0}^{\infty}\alpha_{\mu}^{2}\,e^{-\varepsilon_{\mu}\tau}\,P_{\mu}^{(2a+1,0)}(x)\,P_{\mu}^{(2a+1,0)}(y)\label{kernel}
\end{align}
where $\alpha_{\mu}^{-2}$ is a constant; for Jacobi ensemble, it takes the form
\begin{align}
    \alpha_{\mu}^{-2}=&\frac{2^{2a+1}}{\mu+a+1}.
\end{align}

The integral in Eq.~\eqref{Integraleqn} simplifies considerably when the initial JPDF is chosen as a delta distribution. Physically, the $2\text{CUE} \rightarrow 1\text{CUE}$ crossover is driven by the gradual opening of the junction between the two cavities. At $\tau=0$, the junction is completely closed, implying zero transmission. Consequently, all transmission eigenvalues vanish: $T_{i}=0\Rightarrow y_{i}=1-2T_{i}=1$. This condition translates into the initial JPDF
\begin{align}
    P(\{y\};0)=\prod_{j=1}^{N}\delta(y_{j}-1).\label{initial}
\end{align}
Substituting this in Eq.~\eqref{Integraleqn} requires evaluating a limiting form as $y_{j}\to1$
\begin{align}
    P(\{x\};\tau)=\frac{e^{\mathcal{E}_{0}\tau}}{N!}\Delta_{N}(\{x\})\prod_{i=1}^{N}(1-x_{i})^{2a+1}\lim_{y_{j}\to1}\frac{\text{det}[K_{N}(x_{j},y_{k};\tau)]_{j,k=1\dots N}}{\Delta_{N}(\{y\})},\label{limiteq}
\end{align}
this can be carried out by a L'H\^{o}pital's procedure, the details of which are given in the Supplementary data. After some elementary column operations and simplification, we obtain
\begin{align}
\begin{split}
    P(\{x\};\tau)=\frac{e^{\mathcal{E}_{0}\tau}\,\Delta_{N}(\{x\})\,\prod_{i=1}^{N}(1-x_{i})^{2a+1}}{\prod_{k=1}^{N}\,2^{k-N}\,k\,(k-1)\,\mathcal{B}(k-1,2a+k+1)}\,\det[\hat{h}_{j}(x_{i})]_{i,j=1,\dots,N}\label{eq:finaljpdf}
\end{split}
\end{align}
where $\mathcal{B}(\theta,\phi)$ is the Beta function and
\begin{align}
    \hat{h}_{j}(x)=\frac{j+a}{2^{2a+1}}\,e^{-\varepsilon_{j-1}\tau}\,P_{j-1}^{(2a+1,0)}(x)+\sum_{\nu=N}^{\infty}\frac{\mathcal{K}_{j,\nu}^{(N-j)}}{2^{2a+2}\,\Gamma(2j+2a)}\,e^{-\varepsilon_{\nu}\tau}\,P_{\nu}^{(2a+1,0)}(x)\label{eq:hath}
\end{align}
along with a nontrivial recurrence relation for $\mathcal{K}$
\begin{align}
    \mathcal{K}_{j,\nu}^{(\alpha)}=\mathcal{K}_{j,\nu}^{(\alpha-1)}-\frac{\mathcal{K}_{j,N-\alpha}^{(0)}\,\mathcal{K}_{N-\alpha+1,\nu}^{(\alpha-1)}}{\mathcal{K}_{N-\alpha+1,N-\alpha}^{(0)}}.\label{eq:recur}
\end{align}
Using the structure of $\hat{h}_{j}(x)$ and orthogonality of Jacobi polynomials, the kernel admits a compact form
\begin{align}
    K_{N}(x,y;\tau)=\sum_{\mu=0}^{N-1}\left[(1-x)(1-y)\right]^{(2a+1)/2}\,P_{\mu}^{(2a+1,0)}(x)\,e^{\varepsilon_{\mu}\tau}\,\hat{h}_{\mu+1}(y).\label{eq:finalkernel}
\end{align}

Eqns.~\eqref{eq:finaljpdf}--\eqref{eq:finalkernel} provide a complete description of the Brownian-motion evolution of the Jacobi ensemble relevant to the $2\text{CUE}\to1\text{CUE}$ crossover. In particular, the kernel encapsulates all spectral correlations at finite $\tau$ and serves as the fundamental building block for computing transport observables. In the following section, we use this kernel to obtain the time-dependent statistics of conductance and shot noise power.

\section{\label{sec:level3}Transport in coupled chaotic dots}
Transport properties of mesoscopic systems are obtained using the Landauer-B\"{u}ttiker formalism~\cite{Landauer1996, Fisher1981, Bttiker1990}, by considering electrons as non-interacting quasiparticles tunneling between reservoirs and the mesoscopic region via scattering channels in the leads. This formalism assumes that the transport is phase-coherent and elastic. The two key transport statistics are the Landauer conductance
\begin{align}
    g=&\sum_{i=1}^{N}T_{i},
\end{align}
in units of $e^{2}/h$ and shot noise power
\begin{align}
    p=&\sum_{i=1}^{N}T_{i}(1-T_{i})
\end{align}
in units of $4e^{3}V/h$ where $V$ is a small voltage difference between the two terminals. Shot noise power in mesoscopic conductors captures current fluctuations arising from the discreteness of electric charge~\cite{Bttiker1990, lesovik1989excess}.

Using the kernel derived earlier, we compute the statistics of conductance and shot noise power. By employing the one-point
\begin{align*}
    R_{1}(x;\tau)=K_{N}(x,x;\tau)
\end{align*}
and two-point correlation functions
\begin{align*}
    R_{2}(x,x';\tau)=K_{N}(x,x;\tau)K_{N}(x',x';\tau)-K_{N}(x,x';\tau)K_{N}(x',x;\tau),
\end{align*}
we obtain the average and variance of conductance and shot noise power (the expression for the variance of shot noise power, along with detailed derivations, is provided in the Supplementary data)
\begin{align}
    &\langle g\rangle=\frac{N_{1}\,N_{2}}{N_{\mathrm{s}}}(1-e^{-N_{\mathrm{s}}\tau})\label{avgcond}\\[1.5em]
    \begin{split}
    &\langle p\rangle=\frac{N_{1}^{2}\,N_{2}^{2}}{(N_{\mathrm{s}}-1)\,N_{\mathrm{s}}\,(N_{\mathrm{s}}+1)}+\frac{N_{1}\,N_{2}\,N_{\Delta}^{2}}{(N_{\mathrm{s}}-2)\,N_{\mathrm{s}}\,(N_{\mathrm{s}}+2)}\,e^{-N_{\mathrm{s}}\tau}\\[1.5em]
    &~~~~~~~~+\frac{N_{1}\,N_{2}\,(N_{1}-1)\,(N_{2}-1)}{2(N_{\mathrm{s}}-1)\,(N_{\mathrm{s}}-2)}\,e^{-2(N_{\mathrm{s}}-1)\tau}-\frac{N_{1}\,N_{2}\,(N_{1}+1)\,(N_{2}+1)}{2(N_{\mathrm{s}}+1)\,(N_{\mathrm{s}}+2)}\,e^{-2(N_{\mathrm{s}}+1)\tau}\label{avgshot}
    \end{split}\\[1.2em]
    \begin{split}\label{varg}
    &\mathrm{var}(g)=\frac{N_{1}^{2}\,N_{2}^{2}}{(N_{\mathrm{s}}-1)\,N_{\mathrm{s}}^{2}\,(N_{\mathrm{s}}+1)}+\frac{2N_{1}\,N_{2}\,N_{\Delta}^{2}}{(N_{\mathrm{s}}-2)\,N_{\mathrm{s}}^{2}\,(N_{\mathrm{s}}+2)}\,e^{-N_{\mathrm{s}}\tau}\\[1.5em]
    &~~~~~~~~~~~~+\frac{N_{1}\,N_{2}\,(N_{1}-1)\,(N_{2}-1)}{2(N_{\mathrm{s}}-1)\,(N_{\mathrm{s}}-2)}\,e^{-2(N_{\mathrm{s}}-1)\tau}+\frac{N_{1}\,N_{2}\,(N_{1}+1)\,(N_{2}+1)}{2(N_{\mathrm{s}}+1)\,(N_{\mathrm{s}}+2)}\,e^{-2(N_{\mathrm{s}}+1)\tau}\\[1.5em]
    &~~~~~~~~~~~~-\frac{N_{1}^{2}\,N_{2}^{2}}{N_{\mathrm{s}}^{2}}\,e^{-2N_{\mathrm{s}}\tau}
    \end{split}
\end{align}
with $N_{\mathrm{s}}=N_{1}+N_{2}$. Taking the limit $\tau\to\infty$ (equilibrium), these expressions reproduce transport statistics of a single quantum dot connected with two ideal leads~\cite{Beenakker1997, Savin2006, Sommers2007, Khoruzhenko2009}.

\section{\label{sec:level4}Comparison with numerical simulations}

\subsection{\label{subsec:level4a}Relation between Brownian time and junction width}
To compare our analytical results with numerical simulations of the $2\text{CUE}\rightarrow1\text{CUE}$ crossover, we need to establish a connection between the junction width $n$ and the Brownian time parameter $\tau$. The general interdependence between these variables may be nontrivial, and without any a priori information, its determination can be challenging. The Brownian time $\tau$ is a parameter of the stochastic model and has no automatic relationship to the physical junction width $n$; establishing such a relationship requires either a microscopic derivation (which remains unavailable) or phenomenological mappings. Therefore, we employ heuristic arguments based on elementary probability considerations. These arguments provide a physically motivated starting point for relating the Brownian-time description to the underlying transport geometry.

Following the approach in~\cite{Frahm1995brownian}, we define $p_{\text{out},1}$ as the probability that an electron injected into dot 1 exits back into its own reservoir. Similarly, $p_{\text{trans},1}$ denotes the probability of transmission of this electron into dot 2 through the junction. Analogous definitions hold for the second dot, giving $p_{\text{out},2}$ and $p_{\text{trans},2}$. Using basic probability laws, we arrive at the following relations
\begin{align}
\begin{split}
    &p_{\text{out},1}=\frac{N_{1}}{N_{1}+n},\qquad~~ p_{\text{out},2}=\frac{N_{2}}{N_{2}+n},\\[1.2em]
    &p_{\text{trans},1}=\frac{n}{N_{1}+n},\qquad p_{\text{trans},2}=\frac{n}{N_{2}+n}.
\end{split}
\end{align}

Additionally, we define $p_{\text{total},l}$ as the probability that an electron injected from the left reservoir is ultimately transmitted to the right reservoir through the coupled cavities, and similarly $p_{\text{total},r}$, for injection from the right. Considering all possible paths an electron may take, we arrive at the following self-consistent equations
\begin{align}
\begin{split}
    &p_{l}\equiv p_{\text{total},l}=p_{\text{trans},1}\left(p_{\text{out},2}+p_{\text{trans},2}\,p_{\text{total},l}\right),\\[1.0em]
    &p_{r}\equiv p_{\text{total},r}=p_{\text{trans},2}\left(p_{\text{out},1}+p_{\text{trans},1}\,p_{\text{total},r}\right).
\end{split}
\end{align}

We now use these equations to solve for $p_{l}$ and $p_{r}$, yielding
\begin{align}
    p_{l}=\frac{n\,N_{2}}{N_{1}\,N_{2}+n\,N_{\mathrm{s}}}
\end{align}
and
\begin{align}
    p_{r}=\frac{n\,N_{1}}{N_{1}\,N_{2}+n\,N_{\mathrm{s}}}.
\end{align}
These relations can be useful since, in general, the average conductance can be written as
\begin{gather}
    \langle g\rangle=\frac{N_{1}\,p_{l}+N_{2}\,p_{r}}{2}\label{eq:avgcond}\\[1.2em]
    \Rightarrow\frac{N_{1}\,N_{2}}{N_{\mathrm{s}}}(1-e^{-N_{\mathrm{s}}\tau})=\frac{N_{1}\,p_{l}+N_{2}\,p_{r}}{2}
\end{gather}
where the expression on the left of the latter equation is our analytical result for time-dependent average conductance. Solving the latter equation gives a relation between the time $\tau$ and the junction width $n$
\begin{equation}\label{eq:time&width}
    \tau=\frac{1}{N_{\mathrm{s}}}\ln\left[\frac{N_{1}\,N_{2}+n\,N_{\mathrm{s}}}{N_{1}\,N_{2}}\right].
\end{equation}
We emphasize that the relation above~\eqref{eq:time&width} is obtained by matching the average conductance of the Brownian-motion model to a simple probabilistic description of transport through the coupled cavities. As such, it provides a natural and physically motivated mapping between the junction width $n$ and the Brownian time $\tau$ at the level of ensemble-averaged conductance. However this mapping is phenomenological in nature, since it is constructed by matching ensemble-averaged conductance only. An exact microscopic derivation of the functional dependence between the physical junction width and the Brownian time remains unavailable. Consequently, its validity beyond average conductance, in particular for higher-order transport statistics, is not guaranteed and must be examined independently. To this end, we next construct a random matrix model that directly simulates the $2\text{CUE}\to1\text{CUE}$ crossover of the scattering matrix.

\subsection{\label{subsec:level4b}Matrix Model}

\begin{figure}[!htb]
    \centering
    \includegraphics[width=10cm]{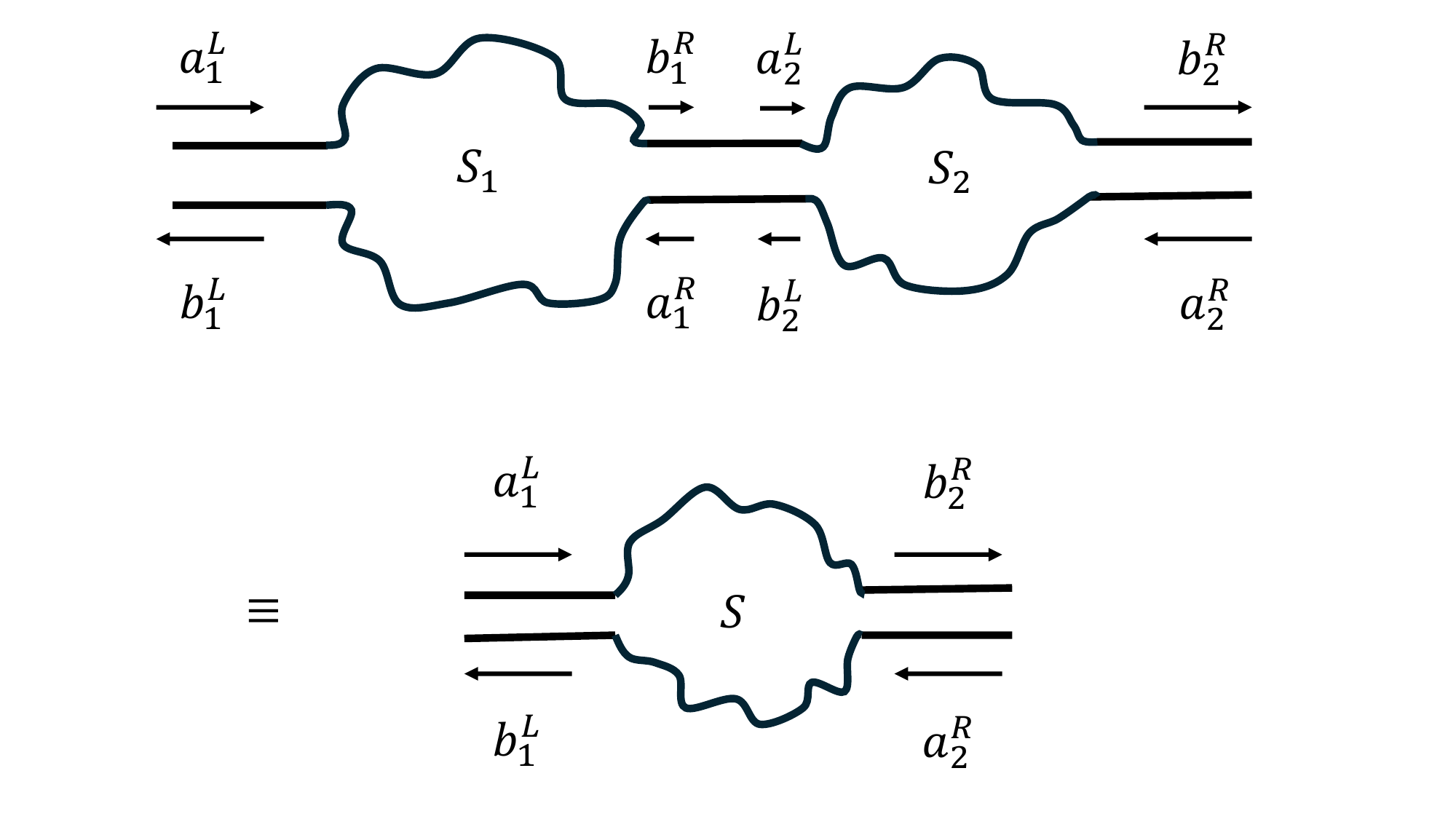}
    \caption{The coupled cavity system can be effectively reduced to a single cavity with the corresponding incoming and outgoing waves.}
    \label{Coupled cavity 2}
\end{figure}

Our model is based on the representation of the two cavities in series using their respective scattering matrices, $S_{1}$ and $S_{2}$, see Fig.~\ref{Coupled cavity 2}. Each cavity connects incoming and outgoing wave amplitudes in its left $(a^{L},\,b^{L})$ and right $(a^{R},\,b^{R})$ leads according to the block structure
\begin{align}
    \begin{pmatrix}
        b_{1}^{L}\\[1.0em]b_{1}^{R}
    \end{pmatrix}=S_{1}\begin{pmatrix}
        a_{1}^{L}\\[1.0em]a_{1}^{R}
    \end{pmatrix}=\begin{pmatrix}
        r_{1} & t_{1}\!'\\[1.0em]t_{1} & r_{1}\!'
    \end{pmatrix}\begin{pmatrix}
        a_{1}^{L}\\[1.0em]a_{1}^{R}
    \end{pmatrix},\label{Sblock1}\\[1.0em]
    \begin{pmatrix}
        b_{2}^{L}\\[1.0em]b_{2}^{R}
    \end{pmatrix}=S_{2}\begin{pmatrix}
        a_{2}^{L}\\[1.0em]a_{2}^{R}
    \end{pmatrix}=\begin{pmatrix}
        r_{2} & t_{2}\!'\\[1.0em]t_{2} & r_{2}\!'
    \end{pmatrix}\begin{pmatrix}
        a_{2}^{L}\\[1.0em]a_{2}^{R}
    \end{pmatrix}.\label{Sblock2}
\end{align}
Since the cavities are connected in series, continuity at the internal lead requires
\begin{align}
\begin{split}
    &b_{1}^{R}=a_{2}^{L},\\[0.8em]
    &a_{1}^{R}=b_{2}^{L}.\label{continuity}
\end{split}
\end{align}

This composite system, however, can equivalently be described by an effective single scattering matrix $S$, as shown in Fig.~\ref{Coupled cavity 2}, with
\begin{align}
    \begin{pmatrix}
        b_{1}^{L}\\[1.0em]b_{2}^{R}
    \end{pmatrix}=S\begin{pmatrix}
        a_{1}^{L}\\[1.0em]a_{2}^{R}
    \end{pmatrix}=\begin{pmatrix}
        r & t'\\[1.0em]t & r'
    \end{pmatrix}\begin{pmatrix}
        a_{1}^{L}\\[1.0em]a_{2}^{R}
    \end{pmatrix}.\label{Sblock}
\end{align}
By combining the relations in Eqs.~\eqref{Sblock1} and~\eqref{Sblock2} with the continuity condition~\eqref{continuity} at the internal lead and comparing with Eq.~\eqref{Sblock}, we can algebraically eliminate the internal amplitudes and express the outgoing amplitudes $(b_{1}^{L},\,b_{2}^{R})$ of the composite scattering matrix solely in terms of its incoming amplitudes $(a_{1}^{L},\,a_{2}^{R})$, thereby relating the composite scattering matrix to the scattering matrices of the individual cavities~\cite{Mello2004}
\begin{align}
\begin{split}
    &r=r_{1}+t_{1}\!'\,(1-r_{2}\,r_{1}\!')^{-1}\,r_{2}\,t_{1}\\[0.8em]
    &t'=t_{1}\!'\,(1-r_{2}\,r_{1}\!')^{-1}\,t_{2}\!'\\[0.8em]
    &t=t_{2}\,(1-r_{1}\!'\,r_{2})^{-1}\,t_{1}\\[0.8em]
    &r'=r_{2}\!'+t_{2}\,(1-r_{1}\!'\,r_{2})^{-1}\,r_{1}\!'\,t_{2}\!'.
\end{split}
\end{align}
These relations can equivalently be obtained using the Redheffer star product~\cite{redheffer1959inequalities, Redheffer1960}, $S=S_{1}\star S_{2}$, which generates the composite scattering matrix.

\begin{figure*}[!t]
    \centering
    \includegraphics[width=0.9\textwidth]{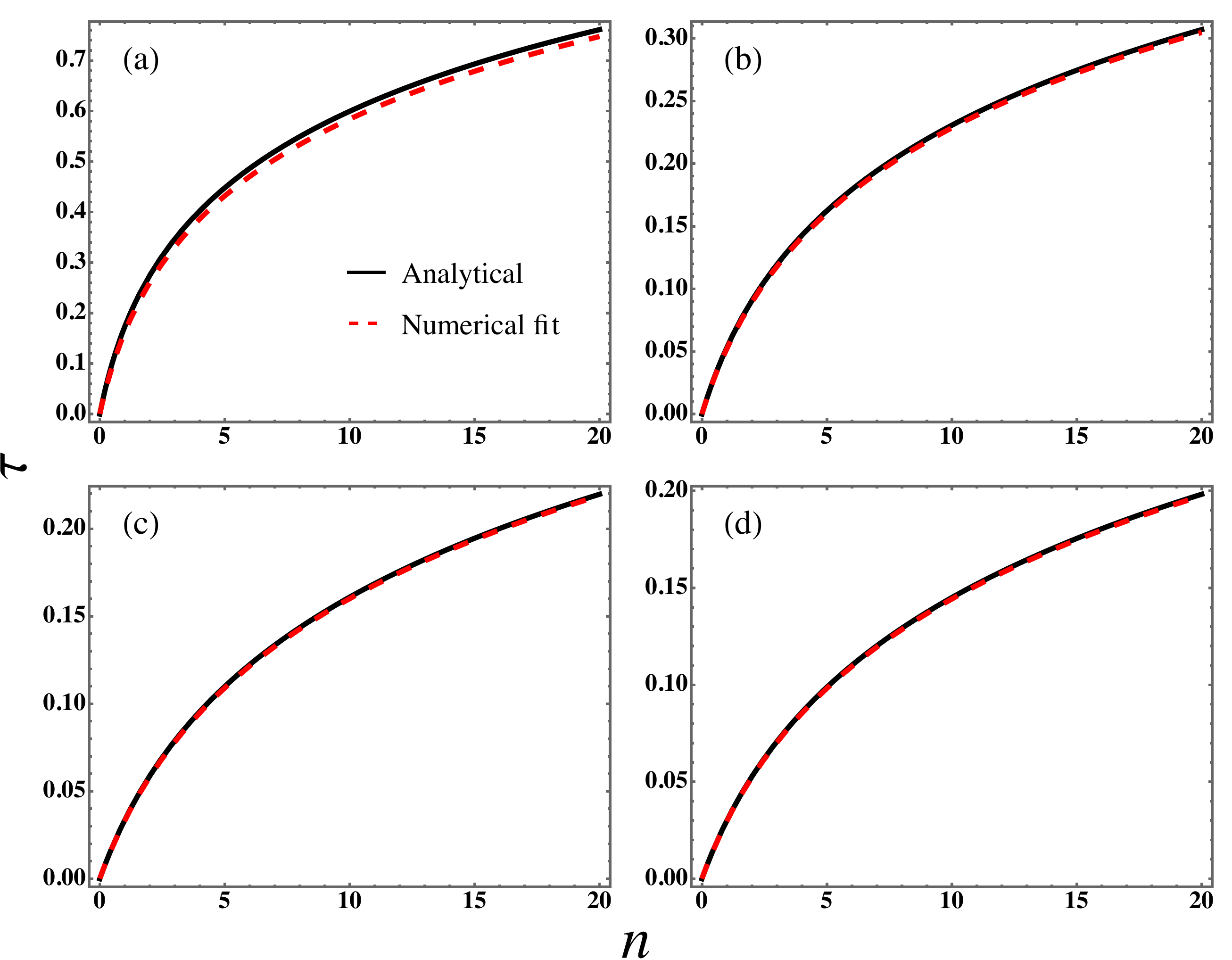}
    \caption{Comparison of the phenomenological $\tau$-$n$ relation~\eqref{eq:time&width} with the numerically fitted $\tau$-$n$ curve obtained from the ensemble-averaged conductance $\langle g\rangle_{M}$, for four lead configurations (a) $N_{1}=N_{2}=2$, (b) $N_{1}=5$, $N_{2}=3$, (c) $N_{1}=5$, $N_{2}=5$, and (d) $N_{1}=7$, $N_{2}=4$, illustrating the accuracy of the phenomenological mapping for conductance.}
    \label{fig:timewithg}
\end{figure*}

We now describe the numerical procedure used to implement the random matrix model and extract transport statistics across the crossover. We perform a simulation of the $2\text{CUE}\to1\text{CUE}$ crossover by generating $S_{1}$ and $S_{2}$ matrices from the CUE, with matrix sizes $(N_{1}+n)\times(N_{1}+n)$ and $(n+N_{2})\times(n+N_{2})$, respectively. For each value of the coupling strength $n$ (varied from 0 to 20 in unit steps), we construct ensembles of $10^{5}$ such random matrices for $(N_{1},N_{2})=(2,2)$, $(5,3)$, $(5,5)$, $(7,4)$. For each realization, we compute the full composite $S$ matrix, extract the transmission block $t'$, and obtain its eigenvalues by diagonalizing $t't'^{\dag}$. The numerically obtained distribution of transmission eigenvalues is then used to evaluate statistics of conductance and shot noise power, namely, $\langle g \rangle_{M}$, $\langle p \rangle_{M}$, $\mathrm{var}(g)_{M}$, and $\mathrm{var}(p)_{M}$ for the matrix model.

We first need to examine the validity of the relation in Eq.~\eqref{eq:time&width} to effectively compare our analytical predictions with numerical simulations. We implement this by numerically determining the value of $\tau$ for each junction width $n$ by matching the analytical conductance~\eqref{avgcond} to the ensemble-averaged matrix model result $\langle g\rangle_{M}$ for the same lead configuration $(N_{1},N_{2})$. This yields a set of $(\tau,n)$ pairs, from which a nonlinear empirical relation is extracted between $\tau$ and $n$, as shown in Fig.~\ref{fig:timewithg}.

\begin{figure*}[!t]
    \centering
    \includegraphics[width=0.9\textwidth]{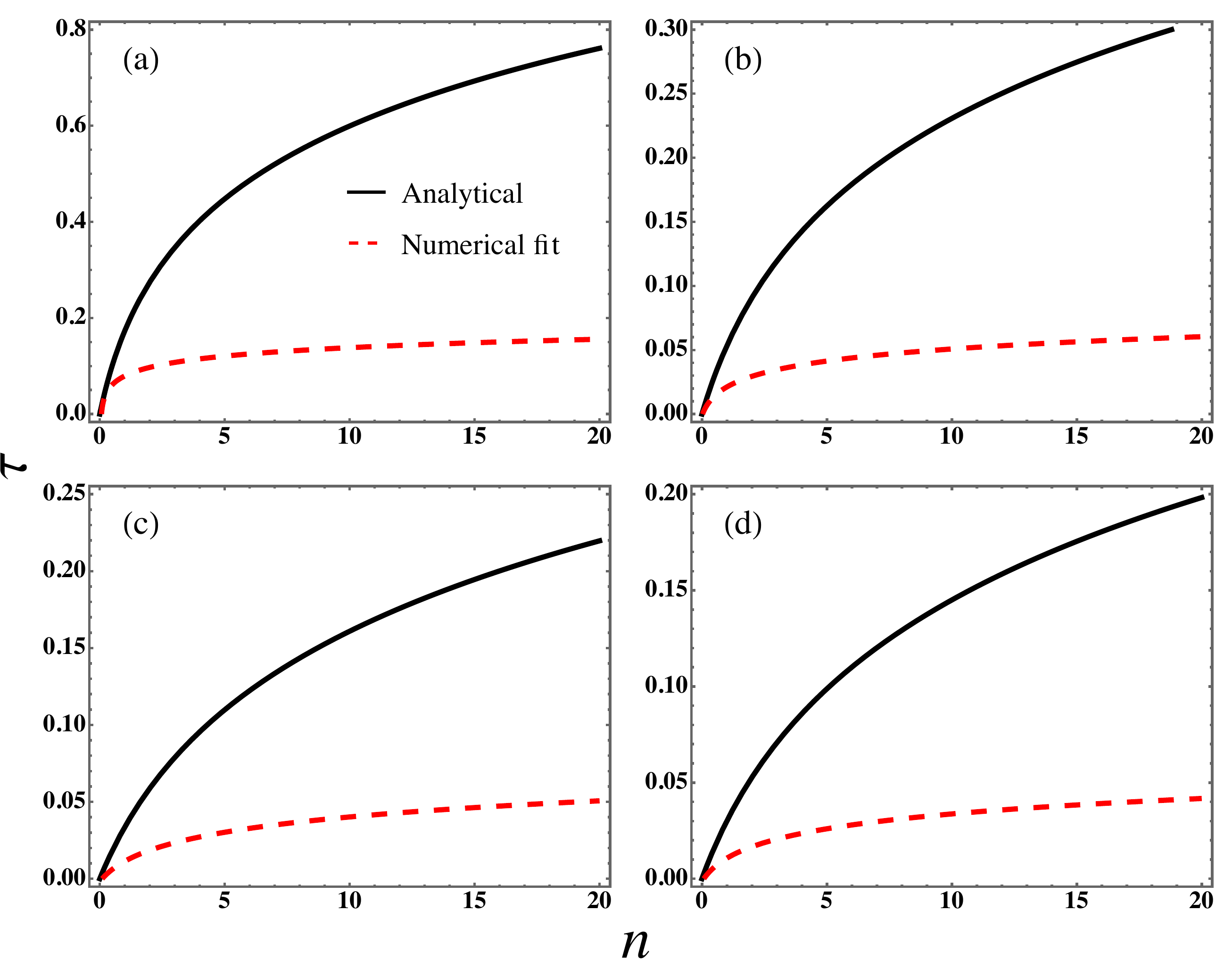}
    \caption{Comparison of the phenomenological $\tau$-$n$ relation~\eqref{eq:time&width} with the numerically fitted $\tau$-$n$ curve obtained from the ensemble-averaged shot noise power $\langle p\rangle_{M}$, for four lead configurations (a) $N_{1}=N_{2}=2$, (b) $N_{1}=5$, $N_{2}=3$, (c) $N_{1}=5$, $N_{2}=5$, and (d) $N_{1}=7$, $N_{2}=4$, this shows that the phenomenological relation~\eqref{eq:time&width} is inadequate for shot noise power, exhibiting clear deviations from the numerically fitted curve. Accordingly, the numerically fitted $\tau$-$n$ relation (dashed curve) extracted from this figure is used in all subsequent analyses involving shot noise power.}
    \label{fig:timewithp}
\end{figure*}

The comparison in Fig.~\ref{fig:timewithg} shows that the analytical relation~\eqref{eq:time&width} provides a good description of the $\tau$-$n$ dependence for the ensemble-averaged conductance. However, repeating the same numerical fitting procedure for the ensemble-averaged shot noise power reveals clear deviations (Fig.~\ref{fig:timewithp}), indicating that Eq.~\eqref{eq:time&width} is inadequate for shot noise power statistics. This is expected because Eq.~\eqref{eq:time&width} is a phenomenological relation constructed by matching the average conductance alone, and its applicability to other observables is therefore not guaranteed. More fundamentally, the observable dependence of the calibration can be understood from the underlying Hamiltonian description of quantum transport. In the Heidelberg formulation, the scattering matrix is given by~\cite{Mahaux1969}
\begin{align}
S(E)=\mathbf{1}-2\pi i W^{\dagger}
\left(E-H+i\pi WW^{\dagger}\right)^{-1}W,
\end{align}
where $H$ is the internal Hamiltonian of the cavity and $W$ represents the coupling between the cavity and leads. Defining the effective non-Hermitian Hamiltonian as $H_{\mathrm{eff}}=H-i\pi WW^{\dagger}$, its right and left eigenvectors satisfy
\begin{align}
H_{\mathrm{eff}}|R_{\mu}\rangle=z_{\mu}|R_{\mu}\rangle,\qquad
\langle L_{\mu}|H_{\mathrm{eff}}=z_{\mu}\langle L_\mu|,
\end{align}
with biorthogonal normalization $\langle L_{\mu}|R_{\nu}\rangle=\delta_{\mu\nu}$. The resolvent can therefore be expanded as
\begin{align}
(E-H_{\mathrm{eff}})^{-1}
=\sum_{\mu}\frac{|R_{\mu}\rangle\langle L_{\mu}|}{E-z_{\mu}}.
\end{align}
Now, $S$ operates in channel space. Therefore, to obtain its matrix element between incoming channel $c'$ and outgoing channel $c$,
\begin{align}
S_{cc'}(E)=\langle c|S(E)|c'\rangle
=\delta_{cc'}-2\pi i\sum_{\mu}
\frac{\langle c|W^{\dagger}|R_{\mu}\rangle
\langle L_{\mu}|W|c'\rangle}{E-z_{\mu}}.
\end{align}
For transmission between different leads, the direct term vanishes, and hence the transmission amplitudes are determined by products of coupling amplitudes,
\begin{align}
t_{cc'}(E)
=-2\pi i\sum_{\mu}
\frac{\langle c|W^{\dagger}|R_{\mu}\rangle
\langle L_{\mu}|W|c'\rangle}{E-z_{\mu}}.
\end{align}
The coupling amplitudes, $\langle c|W^{\dagger}|R_{\mu}\rangle$ and $\langle L_{\mu}|W|c'\rangle$ are projections of the resonance eigenfunctions onto the lead channels, $\{|c\rangle\}$~\cite{Mahaux1969, Poli2009}. Consequently, conductance, which is quadratic in the transmission amplitudes, probes lower-order correlations of these eigenfunction projections, whereas shot-noise power contains the additional term $\operatorname{Tr}[(tt^\dagger)^2]$ and therefore involves higher-order correlations. This provides a microscopic distinction between the eigenfunction measures entering the two transport observables, despite both being expressible in terms of the same transmission eigenvalues. The observable-dependent evolution is then consistent with the Brownian ensemble analysis of eigenfunction statistics, where different eigenfunction measures can exhibit different evolution with the underlying complexity parameter~\cite{Shukla2007}; the connection between transport statistics and eigenfunction measures was further demonstrated for Coulomb-blockade conductance in Ref.~\cite{Dey2011}. Thus, the $\tau$--$n$ relation obtained by calibrating the Brownian model against a particular transport observable should be regarded as an observable-dependent effective calibration, rather than as a universal, geometry-only mapping that would transfer unchanged across observables. Therefore, in all subsequent analyses, we use Eq.~\eqref{eq:time&width} for $\langle g\rangle$, whereas, for shot noise power, we use the numerically fitted $\tau$-$n$ relation extracted from Fig.~\ref{fig:timewithp}.

\subsection{\label{subsec:level4c}Tight-Binding Simulations Using Kwant}

\begin{figure}[!htb]
    \centering
    \includegraphics[width=0.8\textwidth]{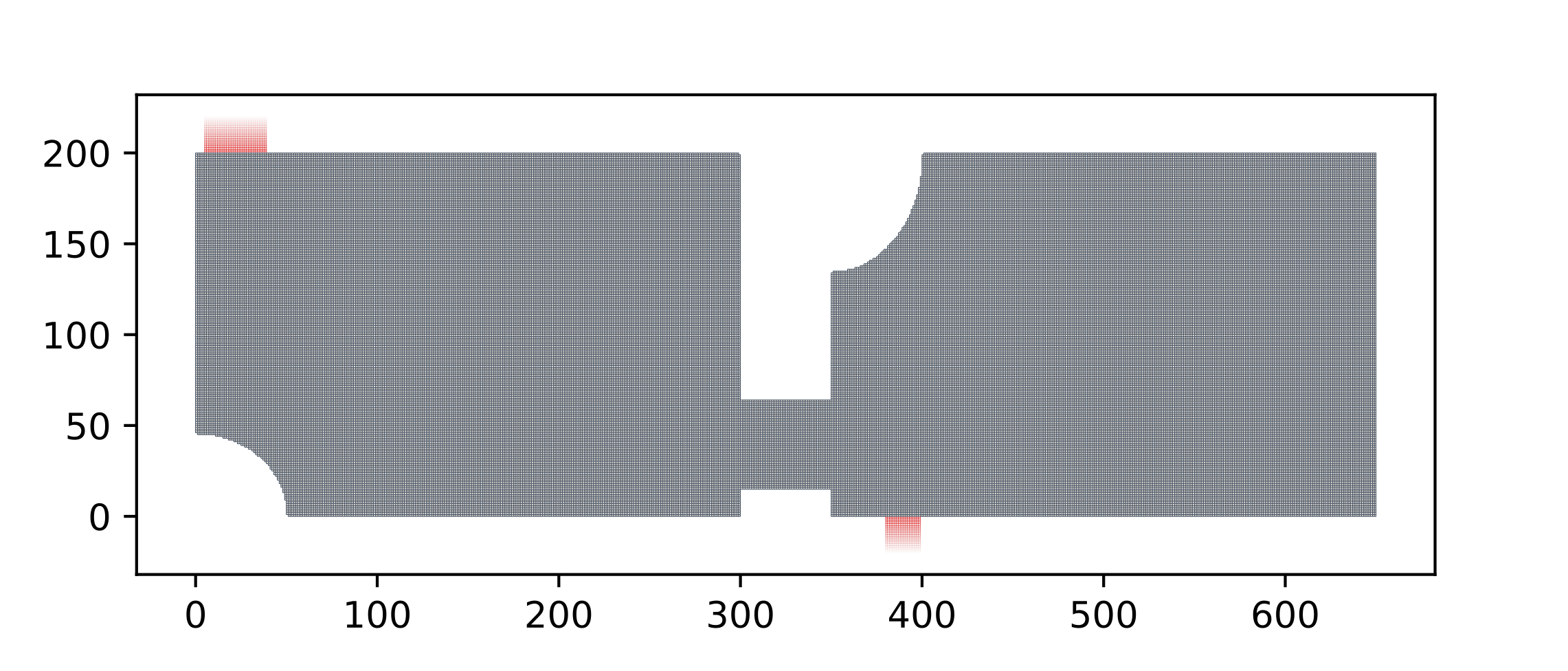}
    \caption{Two-quarter Sinai billiards coupled via a junction of finite width and connected to reservoirs via ideal leads (shown in red).}
    \label{fig:Kwant cavity}
\end{figure}

We also verify our analytical predictions by numerically simulating the coupled cavity system using a tight-binding model implemented in the Kwant framework~\cite{Groth2014}. Kwant is an open-source Python package designed specifically for efficient simulations of tight-binding Hamiltonians and quantum transport calculations. We follow the methodology implemented in~\cite{Chandramouli2020}, with the modifications required for the coupled-cavity geometry considered here.

Our setup consists of a square lattice with lattice constant $a$ which we set to unity. We model the coupled cavity system using two quarter-Sinai billiards of size $300\times 200$, connected in series through a junction whose width $n$ is varied in integer lattice units (see Fig.~\ref{fig:Kwant cavity}). The resulting central region forms the coupled chaotic cavity. Ideal leads are attached to this system by modeling them as semi-infinite translationally invariant lattice systems whose widths can be adjusted in units of $a$~\cite{Groth2014}. The widths of the leads are chosen to realize the desired numbers of open channels $N_{1}$ and $N_{2}$. Sinai billiards are paradigmatic examples of classically chaotic billiards~\cite{sinai1963foundations, Sinai1970}, whose scattering statistics are well described by the circular ensembles of RMT~\cite{Lee1997}. In particular, breaking time-reversal symmetry places the scattering matrix in the unitary symmetry class. The tight-binding Hamiltonian of the system is
\begin{align*}
    H=\epsilon_{0}\sum_{i}c_{i}^\dagger c_{i}-t_{0}\sum_{\langle i,j\rangle}e^{i\phi_{ij}}\,c_{i}^{\dagger}c_{j}+\mathrm{h.c.},
\end{align*}
where $c_{i}^{\dagger}(c_{i})$ creates (annihilates) an electron at site $i$, $\epsilon_{0}=4t_{0}$ is the on-site energy, $t_{0}=\hbar^{2}/2ma^{2}$ is the nearest-neighbour hopping amplitude. A perpendicular magnetic field is introduced through the Peierls substitution~\cite{Groth2014, Peierls1933, Hofstadter1976}, where a phase factor $\exp(i\phi_{ij})=\exp(-\frac{ie}{\hbar c}\int_{\mathbf{r_{1}}}^{\mathbf{r_{2}}}d\mathbf{r}\cdot \mathbf{A(\mathbf{r})})$ is multiplied with the hopping parameter $t_{0}$. $\mathbf{A}$ is the magnetic vector potential and $\mathbf{r_{1}}$, $\mathbf{r_{2}}$ are the lattice sites where hopping is implemented. We work in units $t_{0}=a=e/\hbar c=1$. Choosing the gauge $A_{x}=-By$ and $A_{y}=0$ gives a uniform magnetic field of strength $B$ perpendicular to the plane of the billiard system. In this gauge, hopping along the $x$-direction acquires the phase factor $\exp(iBy)$, while hopping along the $y$-direction remains unchanged. We set $B=1\times10^{-2}$, sufficient to fully break time-reversal symmetry~\cite{Chandramouli2020}.

To drive the crossover, we increase the junction width $n$ in unit steps. For small $n$, the cavities remain effectively decoupled. As the junction widens, electronic coupling between the cavities increases, allowing current to flow. To generate ensembles of scattering matrices, for each value of junction width $n$, we vary the excitation energy from 0.40 to 0.45 in steps of $5\times10^{-6}$, thereby generating an ensemble of 10000 scattering matrices. The energy and lead width determine the number of propagating channels in the system and the dimension of the scattering matrix. The chosen energy range also lies in the low-energy regime where the tight-binding dispersion provides a good approximation to the corresponding continuum dispersion~\cite{Chandramouli2020}. Additionally, we slightly perturb the lead positions every 2,000 iterations to increase ensemble diversity. For each iteration, we compute the conductance and shot noise power using Kwant's built-in functions: smat.transmission and kwant.physics.two\_terminal\_shotnoise(smatrix). The simulations are performed for the same lead configurations considered in the matrix-model analysis.

\begin{figure*}[!t]
    \centering
    \includegraphics[width=0.9\textwidth]{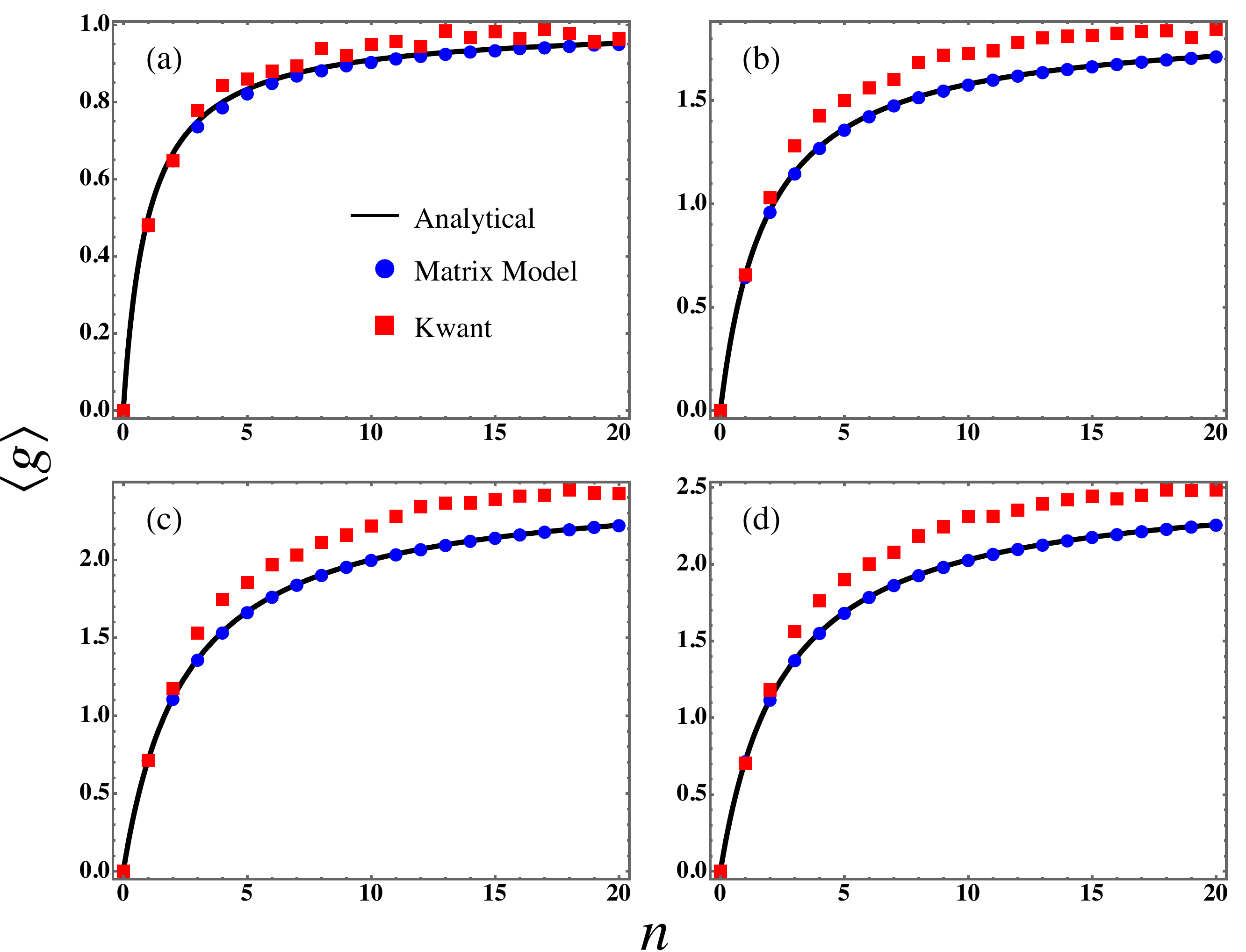}
    \caption{Ensemble-averaged conductance $\langle g\rangle$ as a function of junction width $n$, for (a) $N_{1}=N_{2}=2$, (b) $N_{1}=5$, $N_{2}=3$, (c) $N_{1}=5$, $N_{2}=5$, and (d) $N_{1}=7$, $N_{2}=4$. The analytical curve is obtained by expressing $\tau$ as a function of $n$ in Eq.~\eqref{avgcond} using the phenomenological $\tau$-$n$ relation Eq.~\eqref{eq:time&width}. Agreement between analytical, matrix model, and Kwant results confirms the validity of our analytical expression for ensemble-averaged conductance across the full crossover.}
    \label{fig:avcond}
\end{figure*}

\begin{figure*}[!t]
    \centering
    \includegraphics[width=0.9\textwidth]{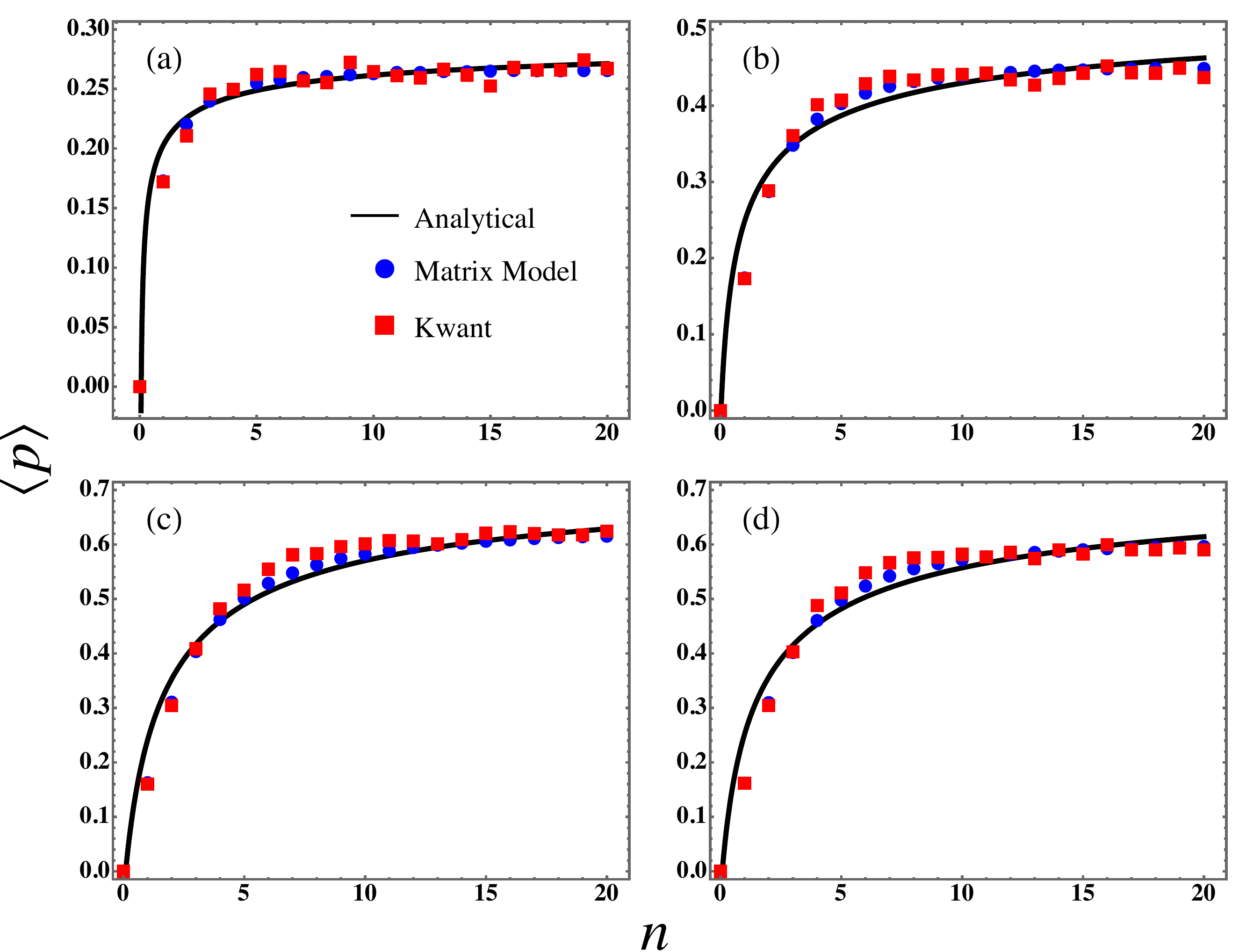}
    \caption{Ensemble-averaged shot noise power $\langle p\rangle$ as a function of junction width $n$, for (a) $N_{1}=N_{2}=2$, (b) $N_{1}=5$, $N_{2}=3$, (c) $N_{1}=5$, $N_{2}=5$, and (d) $N_{1}=7$, $N_{2}=4$. The analytical curve is obtained by expressing $\tau$ as a function of $n$ in Eq.~\eqref{avgshot} using the numerically fitted $\tau$-$n$ relation from Fig.~\ref{fig:timewithp}. Agreement between analytical, matrix model, and Kwant results confirms the validity of our analytical expression for ensemble-averaged shot noise power across the full crossover.}
    \label{fig:avshot}
\end{figure*}

The ensemble-averaged conductance $\langle g \rangle$ and shot noise power $\langle p \rangle$ computed from these simulations are compared against analytical predictions as well as matrix model results in Figs.~\ref{fig:avcond} and~\ref{fig:avshot}. We find good agreement in all cases, confirming the validity of our analytical results in capturing the crossover dynamics at the level of ensemble averages. We emphasize that the Kwant simulations provide a genuinely independent validation of the analytical expressions. Unlike the matrix model comparisons in Figs.~\ref{fig:timewithg} and~\ref{fig:timewithp}, where the $\tau$-$n$ mapping is extracted by fitting the analytical expressions directly to the matrix model data, making agreement with the matrix model partly expected by construction, the Kwant tight-binding simulations are based on a microscopic Hamiltonian with no Dyson Brownian motion structure built in. The $\tau$-$n$ mapping used in Figs.~\ref{fig:avcond} and~\ref{fig:avshot} was derived entirely from the matrix model, and the agreement with Kwant is therefore a non-trivial confirmation that our analytical expressions correctly describe the crossover physics of a realistic chaotic cavity system.

Although the analytical averages of conductance and shot noise power show good agreement with numerical simulations, matching the variances is significantly more challenging. In particular, while the variance of conductance shows reasonable agreement at large $n$, noticeable deviations appear at small $n$. For the variance of shot noise power, however, even qualitative agreement could not be achieved with the current approach. This limitation is not unexpected, it stems from the fact that averages tend to exhibit smoother and more universal behavior across ensemble crossovers, making them amenable to interpolation schemes. In contrast, variances are more sensitive to microscopic fluctuations and system-specific details. In the absence of an exact analytical framework relating $\tau$ and $n$ for higher moments, accurately verifying variances throughout the full crossover remains difficult. Therefore, we restrict the present quantitative comparison to ensemble-averaged observables, which provide the most robust measure of universal transport.

\section{\label{sec:level5}Conclusion}
Dyson’s Brownian motion model provides a powerful framework for describing eigenvalue dynamics in complex quantum systems undergoing symmetry crossovers through matrix-valued diffusion. In this work, we derived closed-form expressions for transport statistics in a system of two chaotic quantum dots coupled via a junction of tunable width and connected to reservoirs with asymmetric ideal leads, when time-reversal symmetry is broken. As the junction width increases, the composite $S$-matrix undergoes a stochastic evolution described by Dyson’s Brownian motion model, exhibiting a crossover from two independent CUE blocks to an effective single CUE.

We derived an explicit closed-form expression for the time-dependent joint probability density of transmission eigenvalues, leading to analytical results for conductance and shot noise power statistics. A key step was the construction of a heuristic mapping between the physical junction width $n$ and the Brownian time parameter $\tau$, which allowed us to connect the crossover in the physical system to its random matrix description. These results were benchmarked against matrix model simulations based on composite scattering matrices, and tight-binding calculations of quarter-Sinai billiards using Kwant. In both cases, we found very good agreement for ensemble-averaged quantities across the entire crossover regime.

While the averages exhibit robust correspondence with numerical data, variances remain more challenging due to their sensitivity to microscopic fluctuations and the lack of an exact analytical relation between $\tau$ and $n$. The present limitation is not one of the model itself, but of establishing a precise $\tau$-$n$ correspondence.

Despite this, the consistent agreement observed at the level of averages demonstrates that the essential physics of the crossover is correctly captured. Our work contributes to a broader understanding of transport in mesoscopic systems where scattering regions can be coupled continuously. Extensions to other symmetry crossovers constitute natural directions for future investigation. The determinantal structure of the time-dependent joint probability density in Eq.~\eqref{eq:finaljpdf} further suggests that Andr\'{e}ief-type integration identities~\cite{andreief1883note} could, in principle, be employed to derive full time-dependent probability distributions of conductance and shot noise power. We leave this as a promising direction for future investigation.

\appendix

\section{Solution of the Fokker-Planck equation via a limiting procedure}

We employ L'H\^{o}pital's rule to evaluate the limit
\begin{align}
    P(\{x\};\tau)=\frac{e^{\mathcal{E}_{0}\tau}}{N!}\,\Delta_{N}(\{x\})\prod_{i=1}^{N}(1-x_{i})^{2a+1}\lim_{y_{j}\to1}\,\frac{\det\left[K_{N}(x_{j},y_{k};\tau)\right]_{j,k=1,\dots,N}}{\Delta_{N}(\{y\})}\label{limitjpd}
\end{align}
following the approach outlined in Appendix B of~\cite{Pivaro2017}. This involves expressing the numerator as a determinant matrix whose $(i,j)$th element is given by
\begin{align*}
    \frac{d^{j-1}}{dy_{j}^{j-1}}K_{N}(x_{i},y_{j};\tau)\bigg|_{y_{j}=1}
    =\sum_{\mu=0}^{\infty}\frac{2\mu+2a+2}{2^{2a+2}}\,e^{-\varepsilon_{\mu}\tau}\,P_{\mu}^{(2a+1,\,0)}(x_{i})\,\frac{d^{j-1}}{dy_{j}^{j-1}}\,P_{\mu-j+1}^{(2a+j,\,j-1)}(y_{j})\bigg|_{y_{j}=1}
\end{align*}
valid for $\mu+1>j-1$ (since $P_{\mu-j+1}^{(2a+j,\,j-1)}$ is defined only for $\mu-j+1\geq 0$, i.e. $\mu\geq j-1$). Using the identity for the derivative of the Jacobi polynomials
\begin{align*}
    \frac{d^{r}}{dw^{r}}P_{\mu}^{(u,v)}(w)=\frac{\Gamma(u+v+\mu+r+1)}{2^{r}\,\Gamma(u+v+\mu+1)}\,P_{\mu-r}^{(u+r,v+r)}(w).
\end{align*}
along with the series expansion
\begin{align*}
    P_{\mu}^{(u,v)}(w)=\frac{\Gamma(u+\mu+1)}{\mu!\,\Gamma(u+v+\mu+1)}\,\sum_{m=0}^{\mu}\,\binom{\mu}{m}\,\frac{\Gamma(u+v+\mu+m+1)}{\Gamma(u+m+1)}\,\left(\frac{w-1}{2}\right)^{m},
\end{align*}
the $(i,j)$th element of the determinant in the numerator simplifies to
\begin{align}
\begin{split}
    &\frac{d^{j-1}}{dy_{j}^{j-1}}K_{N}(x_{i},y_{j};\tau)\bigg|_{y_{j}=1}=\\\\
    &\begin{cases}
        \displaystyle
        \frac{1}{2^{2a+1}}\sum_{\mu=j-1}^{\infty}\frac{\mu+a+1}{2^{j-1}}\frac{\Gamma(\mu+2a+j+1)}{\Gamma(\mu-j+2)\,\Gamma(2a+j+1)}e^{-\varepsilon_{\mu}\tau}\,P_{\mu}^{(2a+1,0)}(x_{i}), & \text{if } j-1<\mu+1,\\[2em]
        0, & \text{otherwise.}
    \end{cases}
\end{split}
\end{align}

Applying the same procedure to the denominator, the determinant matrix in this case reduces to a product of gamma functions $\prod_{k=1}^{N}\Gamma(k+1)$. Hence, the JPDF in~\eqref{limitjpd} reduces to the following form after simplification
\begin{align}
    P(\{x\};\tau)=\frac{e^{\mathcal{E}_{0}\tau}}{2^{N\left(2a+1\right)}}\,\frac{\Delta_{N}(\{x\})\,\prod_{i=1}^{N}\,(1-x_{i})^{2a+1}}{\prod_{k=1}^{N}\,2^{k}\,\Gamma(k+1)\,\Gamma(2a+k+1)}\,\det[h_{j}(x_{i})]_{i,j=1,\dots,N}
\end{align}
where
\begin{align}
    h_{j}(x)=\sum_{\mu=j-1}^{\infty}\frac{(2\mu+2a+2)\,\Gamma(2a+\mu+j+1)}{\Gamma(\mu-j+2)}\,e^{-\varepsilon_{\mu}\tau}\,P_{\mu}^{(2a+1,0)}(x).
\end{align}

Let us define the coefficient
\begin{align*}
    \mathcal{K}_{j,\,\mu}^{(0)}=\frac{(2\mu+2a+2)\,\Gamma(2a+\mu+j+1)}{\Gamma(\mu-j+2)},
\end{align*}
treating $h_{j}(x)$ as the $j$th column vector $c_{j}$ of the matrix $h(x)$, we perform the column operation
\begin{align}
    c_{j} \;\;\rightarrow\;\; c_{j}\,' = 
    \begin{cases}
        \displaystyle 
        c_{j} - 
        \dfrac{\mathcal{K}_{j,\,N-1}^{(0)}}
              {\mathcal{K}_{N,\,N-1}^{(0)}}\,c_{N},
        & j=1,\dots,N-1, \\[1.5em]
        c_{N}, & j = N.
    \end{cases}
    \label{recurrence}
\end{align}
where
\begin{align}
    c_{N}=h_{N}(x)=\sum_{\mu=N-1}^{\infty}\mathcal{K}_{N,\,\mu}^{(0)}\,e^{-\varepsilon_{\mu}\tau}\,P_{\mu}^{(2a+1,0)}(x).\label{eq:hcolumn}
\end{align}
This gives
\begin{align*}
    c_{j}\,'=\sum_{\mu=j-1}^{\infty}\mathcal{K}_{j,\,\mu}^{(0)}\,e^{-\varepsilon_{\mu}\tau}\,P_{\mu}^{(2a+1,0)}
    (x)-\sum_{\mu=N-1}^{\infty}\frac{\mathcal{K}_{j,N-1}^{(0)}\,\mathcal{K}_{N,\mu}^{(0)}}{\mathcal{K}_{N,N-1}^{(0)}}\,e^{-\varepsilon_{\mu}\tau}\,P_{\mu}^{(2a+1,0)}(x)
\end{align*}
for all $j=1,\dots,N-1$. We split the above two sums separately and find that the $\mu=N-1$ term cancels out, giving
\begin{align}
    c_{j}\,'=h_{j}(x)=\sum_{\mu=j-1}^{N-2}\mathcal{K}_{j,\mu}^{(0)}\,e^{-\varepsilon_{\mu}\tau}\,P_{\mu}^{(2a+1,0)}(x)+\sum_{\mu=N}^{\infty}\left(\mathcal{K}_{j,\mu}^{(0)}-\frac{\mathcal{K}_{j,N-1}^{(0)}\,\mathcal{K}_{N,\mu}^{(0)}}{\mathcal{K}_{N,N-1}^{(0)}}\right)\,e^{-\varepsilon_{\mu}\tau}\,P_{\mu}^{(2a+1,0)}(x).
\end{align}
Now we can again consider the term in the braces above as $\mathcal{K}_{j,\mu}^{(1)}$ and perform similar column operations as in~\eqref{recurrence} till only the first term corresponding to $\mu=j-1$ remains in the first sum. We end up with
\begin{align}
    h_{j}(x)=\mathcal{K}_{j,j-1}^{(0)}\,e^{-\varepsilon_{j-1}\tau}\,P_{j-1}^{(2a+1,0)}(x)+\sum_{\mu=N}^{\infty}\,\mathcal{K}_{j,\mu}^{(N-j)}\,e^{-\varepsilon_{\mu}\tau}\,P_{\mu}^{(2a+1,0)}(x)
\end{align}
and a recurrence relation for $\mathcal{K}$
\begin{equation}
    \mathcal{K}_{j,\mu}^{(\alpha)}=\mathcal{K}_{j,\mu}^{(\alpha-1)}-\frac{\mathcal{K}_{j,N-\alpha}^{(0)}\,\mathcal{K}_{N-\alpha+1,\mu}^{(\alpha-1)}}{\mathcal{K}_{N-\alpha+1,N-\alpha}^{(0)}}.\label{eq:recurrentK}
\end{equation}
We normalize $h(x)$ to ensure proper normalization in the correlation kernel construction
\begin{align}
    \hat{h}_{j}(x)=\frac{2j+2a}{2^{2a+2}}\,e^{-\varepsilon_{j-1}\tau}\,P_{j-1}^{(2a+1,0)}(x)+\sum_{\nu=N}^{\infty}\,\frac{\mathcal{K}_{j,\nu}^{(N-j)}}{2^{2a+2}\,\Gamma(2j+2a)}\,e^{-\varepsilon_{\nu}\tau}\,P_{\nu}^{(2a+1,0)}(x),
\end{align}
Here we have relabeled the summation index in the second term from $\mu$ to $\nu$, so that when $\hat{h}_{j}$ appears inside the kernel $K_{N}$ in Eq.~\eqref{eq:finalkernel} of the main text, the inner summation index $\nu$ is distinct from the outer index $\mu$. Note that in Eq.~\eqref{eq:recurrentK} the $\mathcal{K}$ coefficients carry index $\mu$; in the corresponding Eq.~\eqref{eq:recur} of the main text they carry index $\nu$, consistent with the relabelling above. The kernel can then be constructed as
\begin{align}
    K_{N}(x,y;\tau)=\sum_{\mu=0}^{N-1}\left[(1-x)(1-y)\right]^{(2a+1)/2}\,P_{\mu}^{(2a+1,0)}(x)\,e^{\varepsilon_{\mu}\tau}\,\hat{h}_{\mu+1}(y)
\end{align}
and
\begin{align}
    P(\{x\};\tau)=\frac{e^{\mathcal{E}_{0}\tau}\,\Delta_{N}(\{x\})\,\prod_{i=1}^{N}\,(1-x_{i})^{2a+1}}{\prod_{k=1}^{N}\,2^{k-N}\,k\,(k-1)\,\mathcal{B}(k-1,\,2a+k+1)}\,\det[\hat{h}_{j}(x_{i})]_{i,j=1,\dots,N}.
\end{align}

\section{Derivation for the transport statistics}

Transport properties are linear statistics on the transmission eigenvalues. For a transport observable of the form
\begin{align}
    \phi=\sum_{i=1}^{N}f(T_{i})
\end{align}
where $f(T)$ is some function of the transmission eigenvalue $T$, its average at time $\tau$ can be computed with the JPDF $P(\{T\};\tau)$
\begin{align}
    \langle \phi\rangle=&\int_{0}^{1}\dots\int_{0}^{1}\phi\,P(\{T\};\tau)\,dT_{1}\dots dT_{N}\\[1.2ex]
    =&\int_{0}^{1}\dots\int_{0}^{1}\sum_{i=1}^{N}f(T_{i})\,P(\{T\};\tau)\,dT_{1}\dots dT_{N}\\[1.2ex]
    =&\sum_{i=1}^{N}\int_{0}^{1}\dots\int_{0}^{1}f(T_{i})\,P(\{T\};\tau)\,dT_{1}\dots dT_{N}.
\end{align}
In the case of linear statistics, $f$ is a symmetric function of the transmission eigenvalues, hence
\begin{align}
    \langle \phi\rangle=&\,N\int_{0}^{1}\dots\int_{0}^{1}f(T_{1})\,P(\{T\};\tau)\,dT_{1}\dots dT_{N}\\
    =&\int_{0}^{1}f(T)\,R_{1}(T;\tau)\,dT=\int_{-1}^{1}f(x)\,K_{N}(x,x;\tau)dx\label{eq:Jacobiavg}
\end{align}
where $R_{1}(T;\tau)$ is the one-point correlation function, and the last line uses the change of variables $x=1-2T$. Likewise, for the variance, we can formulate
\begin{align}
    \text{var}(\phi)=&\langle \phi^{2}\rangle-\langle \phi\rangle^{2}=\int_{0}^{1}\int_{0}^{1}f(T)\,f(T')\,R_{2}(T,T';\tau)\,dTdT'-\langle \phi\rangle^{2}\\[1.2ex]
    \begin{split}
        =&\int_{-1}^{1}\int_{-1}^{1}f\bigg(\frac{1-x}{2}\bigg)\,f\bigg(\frac{1-x'}{2}\bigg)\Big(K_{N}(x,x;\tau)\,K_{N}(x',x';\tau)\\[1.2ex]&~~~-K_{N}(x,x';\tau)\,K_{N}(x',x;\tau)\Big)\,dx\,dx'-\langle \phi\rangle^{2}.
    \end{split}\label{eq:Jacobivar}
\end{align}
For average and variance of conductance and shot noise power, the function $f$ reduces to low-degree polynomials, leading to integrals involving positive powers of $x$ multiplied with the kernel $K_{N}(x,x';\tau)$. These integrals can be solved using the recurrence relation
\begin{align}
\begin{split}
    w\,P_{\mu}^{(u,v)}(w)=\,&\frac{2(\mu+u)(\mu+v)(2\mu+u+v+2)}{(2\mu+u+v)_{3}}P_{\mu-1}^{(u,v)}(w)\\[1.5ex]
    &-\frac{(2\mu+u+v+1)(u^{2}-v^{2})}{(2\mu+u+v)_{3}}P_{\mu}^{(u,v)}(w)\\[1.5ex]
    &+\frac{2(\mu+1)(\mu+u+v+1)(2\mu+u+v)}{(2\mu+u+v)_{3}}P_{\mu+1}^{(u,v)}(w)\label{eq:recurrence}
\end{split}
\end{align}
and the orthogonality relation for the Jacobi polynomials
\begin{align}
\begin{split}
    \int_{-1}^{1}(1-w)^{u}\,(1+w)^{v}\,P_{m}^{(u,v)}(w)\,&P_{\mu}^{(u,v)}(w)\,dw=\\
    &\frac{2^{u+v+1}}{2\mu+u+v+1}\frac{\Gamma(\mu+u+1)~\Gamma(\mu+v+1)}{\Gamma(\mu+u+v+1)~\mu!}~\delta_{\mu,m}\label{eq:orthogonality}
\end{split}
\end{align}
for $u,v>-1$, where $(a)_{\mu}=\Gamma(a+\mu)/\Gamma(a)$ denotes the Pochhammer symbol. 

We now illustrate how these tools yield the closed-form results. For the ensemble-averaged conductance, we set $f(T)=T$, so $f(x)=(1-x)/2$ under the substitution $x=1-2T$. Using Eq.~\eqref{eq:Jacobiavg},
\begin{align*}
    \langle g\rangle=\int_{-1}^{1}\frac{1-x}{2}\, K_N(x,x;\tau)\,dx.
\end{align*}
This can be split into two integrals, the first integral evaluates to $N/2$ by the normalization of the one-point kernel. For the second integral, we substitute the kernel~\eqref{eq:finalkernel} along with the definition of $\hat{h}_j(x)$ from Eq.~\eqref{eq:hath} to obtain
\begin{align}
    \langle g\rangle=\frac{N}{2}-\frac{1}{2}(\mathcal{I}+\mathcal{II})
\end{align}
where
\begin{align}
\mathcal{I}
&= \int_{-1}^{+1} x\,\sum_{\mu=0}^{N-1}
   \frac{2\mu+2a+2}{2^{2a+2}}\,(1-x)^{2a+1}
   \left[P_\mu^{(2a+1,0)}(x)\right]^2 dx, \label{eq:I_def}\\[6pt]
\mathcal{II}
&= \int_{-1}^{+1} x\,\sum_{\mu=0}^{N-1}\sum_{k=N}^{\infty}
   \frac{K^{(N-(\mu+1))}_{\mu+1,\,k}}{2^{2a+2}\,\Gamma(2\mu+2a+2)}\,
   e^{(\varepsilon_\mu - \varepsilon_k)\tau}\,
   (1-x)^{2a+1}\,P_\mu^{(2a+1,0)}(x)\,
   P_k^{(2a+1,0)}(x)\, dx. \label{eq:II_def}
\end{align}
We now apply the three-term recurrence~\eqref{eq:recurrence} to express $xP_{\mu}^{(2a+1,0)}(x)$ as a linear combination of $P_{\mu-1}^{(2a+1,0)}$, $P_{\mu}^{(2a+1,0)}$ and $P_{\mu+1}^{(2a+1,0)}$. For the case of $\mathcal{I}$, integrating against
$(1-x)^{2a+1} P_\mu^{(2a+1,0)}(x)$, only the middle term survives
by the orthogonality relation~\eqref{eq:orthogonality}, which upon simplification gives
\begin{align}
\mathcal{I}=-\frac{N_{2}(N_{1}-N_{2})}{N_{\mathrm{s}}}.
\label{eq:I_result}
\end{align}

For $\mathcal{II}$, we again apply the recurrence~\eqref{eq:recurrence}
to expand $x\,P_\mu^{(2a+1,0)}(x)$ and integrate against $(1-x)^{2a+1} P_k^{(2a+1,0)}(x)$
using orthogonality. The three terms in the recurrence produce integrals
proportional to $\delta_{\mu-1,\,k}$, $\delta_{\mu,\,k}$, and $\delta_{\mu+1,\,k}$
respectively. Since the sum over $k$ runs from $N$ to $\infty$ and the sum over $\mu$ runs from $0$ to $N-1$ in~\eqref{eq:II_def}, the first two terms vanish while the only surviving contribution comes from the third ($\delta_{\mu+1,k}$) term for $\mu=N-1$, $k=N$. Simplifying, we obtain
\begin{align}
\mathcal{II}=\frac{2N_{1}N_{2}}{N_{\mathrm{s}}}\, e^{-N_{\mathrm{s}}\tau}.
\label{eq:II_result}
\end{align}
This leads to the ensemble-averaged conductance
\begin{align}
\langle g\rangle=\frac{N_{1}N_{2}}{N_{\mathrm{s}}}\left(1-e^{-N_{\mathrm{s}}\tau}\right),
\end{align}
which is Eq.~\eqref{avgcond}. The average shot noise power, as well as variance of conductance and shot noise power follows analogously, generating higher-degree polynomials that require additional applications of the recurrence and orthogonality relations.

\section{Variance of shot noise power}

The variance of shot noise power is given by
\begin{align}
\begin{split}
  \mathrm{var}(p)=A_{0}+\sum_{j=1}^{4}\Big[&A_{j,0}e^{-jN_{\mathrm{s}}\tau}+\sum_{k=1}^{j-1}\Big(A_{j,k}e^{-j(N_{\mathrm{s}}+k)\tau}+A_{j,-k}e^{-j(N_{\mathrm{s}}-k)\tau}\Big)\Big]\\[1.5ex]
  &+B_{-}e^{-N_{\mathrm{s}}\tau}e^{-2(N_{\mathrm{s}}-1)\tau}+B_{+}e^{-N_{\mathrm{s}}\tau}e^{-2(N_{\mathrm{s}}+1)\tau}\label{eq:varp}
\end{split}
\end{align}
where the coefficients $A_{0}$, $A_{j,k}$ and $B_{\pm}$ are rational functions of $N_{1}$ and $N_{2}$. The first coefficient takes the form

\begin{align}
\begin{split}
    A_{0}=\,&\frac{N_{1}^{2}\,N_{2}^{2}\left(N_{1}-1\right)^{2}\left(N_{2}-1\right)^{2}}{\left(N_{\mathrm{s}}-3\right)\left(N_{\mathrm{s}}-2\right)^{2}\left(N_{\mathrm{s}}-1\right)^{2}N_{\mathrm{s}}^{2}\left(N_{\mathrm{s}}+1\right)}
    +\frac{N_{1}^{2}\,N_{2}^{2}\left(N_{1}+1\right)^{2}\left(N_{2}+1\right)^{2}}{\left(N_{\mathrm{s}}-1\right)N_{\mathrm{s}}^{2}\left(N_{\mathrm{s}}+1\right)^{2}\left(N_{\mathrm{s}}+2\right)^{2}\left(N_{\mathrm{s}}+3\right)}\\[1.5ex]
    &~~~~~~~+\frac{N_{1}^{2}\,N_{2}^{2}\,N_{\Delta}^{4}}{\left(N_{\mathrm{s}}-2\right)^{2}\left(N_{\mathrm{s}}-1\right)N_{\mathrm{s}}^{2}\left(N_{\mathrm{s}}+1\right)\left(N_{\mathrm{s}}+2\right)^{2}}.
\end{split}
\end{align}

In the limit $\tau\to\infty$, all exponential terms vanish and $\mathrm{var}(p)$ reduces to $A_{0}$, which coincides with the variance of shot noise power for a two–terminal chaotic cavity in the absence of time-reversal symmetry~\cite{Savin2006, Sommers2007, Khoruzhenko2009}. The various coefficients $A_{j,k}$ and $B_{\pm}$ admit a common structural form, and their expressions can be compactly written as
\begin{align}
  A_{j,k}=\frac{Q_{j,k}}{\prod_{m}(N_{\mathrm{s}}+m)^{\alpha_{j,k}(m)}},\qquad B_{\pm}=\frac{R_{\pm}}{\prod_{m}(N_{\mathrm{s}}+m)^{\gamma_{\pm}(m)}},\label{eq:struct}
\end{align}
with $Q_{j,k}$, $R_{\pm}$ polynomials in $N_{1}$ and $N_{2}$ and integer exponents $\alpha_{j,k}(m)$, $\gamma_{\pm}(m)$. Although the representation in~\eqref{eq:struct} is compact, the ranges of $m$ and the exponents $\alpha_{j,k}(m)$, $\gamma_{\pm}(m)$ differ from case to case. For this reason we list their fully expanded forms below. In presenting them, we factor out simple multiplicative terms involving $N_{1}$, $N_{2}$ and $N_{\Delta}$ from the polynomials $Q_{j,k}$ and $R_{\pm}$ in~\eqref{eq:struct}. This does not alter their structural form; it merely isolates common prefactors so that the expressions for $Q_{j,k}$ and $R_{\pm}$ remain readable, especially in the cases where the underlying polynomials are very large.

\noindent
\begin{align*}
A_{1,0}=\frac{2\,N_{1}N_{2}\,N_{\Delta}^{2}}{(N_{\mathrm{s}}-4)(N_{\mathrm{s}}-3)(N_{\mathrm{s}}-2)(N_{\mathrm{s}}-1)\,N_{\mathrm{s}}^{2}(N_{\mathrm{s}}+1)(N_{\mathrm{s}}+2)(N_{\mathrm{s}}+3)(N_{\mathrm{s}}+4)}\,Q_{1,0},
\end{align*}
\begin{align*}
A_{2,-1}=\frac{N_{1}\,N_{2}\,(N_{1}-1)\,(N_{2}-1)}{2\,(N_{\mathrm{s}}-5)\,(N_{\mathrm{s}}-4)\,(N_{\mathrm{s}}-2)\,(N_{\mathrm{s}}-1)^{2}\,N_{\mathrm{s}}(N_{\mathrm{s}}+2)\,(N_{\mathrm{s}}+3)}\,Q_{2,-1},
\end{align*}
\begin{align*}
A_{2,0}=\frac{N_{1}^{2}\,N_{2}^{2}\,N_{\Delta}^{4}}{(N_{\mathrm{s}}-2)^{2}\,N_{\mathrm{s}}^{2}\,(N_{\mathrm{s}}+2)^{2}}\,Q_{2,0},
\end{align*}
\begin{align*}
A_{2,1}=\frac{N_{1}\,N_{2}\,(N_{1}+1)\,(N_{2}+1)}{2\,(N_{\mathrm{s}}-3)\,(N_{\mathrm{s}}-2)\,N_{\mathrm{s}}\,(N_{\mathrm{s}}+1)^{2}\,(N_{\mathrm{s}}+2)\,(N_{\mathrm{s}}+4)\,(N_{\mathrm{s}}+5)}\,Q_{2,1},
\end{align*}
\begin{align*}
A_{3,-2}=\frac{N_{1}\,N_{2}\,(N_{1}-1)\,(N_{2}-1)\,(N_{1}-2)\,(N_{2}-2)\,N_{\Delta}^{2}}{3\,(N_{\mathrm{s}}-6)\,(N_{\mathrm{s}}-4)\,(N_{\mathrm{s}}-3)\,(N_{\mathrm{s}}-2)\,(N_{\mathrm{s}}+2)}\,Q_{3,-2},
\end{align*}
\begin{align*}
A_{3,0}=-\frac{8N_{1}\,N_{2}\,(N_{1}^{2}-1)\,(N_{2}^{2}-1)\,N_{\Delta}^{2}}{3\,(N_{\mathrm{s}}-4)\,(N_{\mathrm{s}}-2)\,N_{\mathrm{s}}^{2}\,(N_{\mathrm{s}}+2)\,(N_{\mathrm{s}}+4)}\,Q_{3,0},
\end{align*}
\begin{align*}
A_{3,2}=-\frac{N_{1}\,N_{2}\,(N_{1}+1)\,(N_{2}+1)\,(N_{1}+2)\,(N_{2}+2)\,N_{\Delta}^{2}}{3\,(N_{\mathrm{s}}-2)\,(N_{\mathrm{s}}+2)\,(N_{\mathrm{s}}+3)\,(N_{\mathrm{s}}+4)\,(N_{\mathrm{s}}+6)}\,Q_{3,2},
\end{align*}
\begin{align*}
A_{4,-3}=\frac{N_{1}\,N_{2}\,(N_{1}-1)\,(N_{2}-1)\,(N_{1}-2)\,(N_{2}-2)\,(N_{1}-3)\,(N_{2}-3)}{24\,(N_{\mathrm{s}}-6)\,(N_{\mathrm{s}}-5)\,(N_{\mathrm{s}}-4)\,(N_{\mathrm{s}}-3)}\,Q_{4,-3},
\end{align*}
\begin{align*}
A_{4,-1}=\frac{N_{1}\,N_{2}\,(N_{1}-1)\,(N_{2}-1)}{8\,(N_{\mathrm{s}}-1)^{2}}\,Q_{4,-1},
\end{align*}
\begin{align*}
A_{4,0}=\frac{2N_{1}^{2}\,N_{2}^{2}\,(N_{1}^{2}-1)\,(N_{2}^{2}-1)}{3\,(N_{\mathrm{s}}-2)\,N_{\mathrm{s}}^{2}\,(N_{\mathrm{s}}+2)}\,Q_{4,0},
\end{align*}
\begin{align*}
A_{4,1}=\frac{N_{1}\,N_{2}\,(N_{1}+1)\,(N_{2}+1)}{8(N_{\mathrm{s}}+1)^{2}}\,Q_{4,1},
\end{align*}
\begin{align*}
A_{4,3}=\frac{N_{1}\,N_{2}\,(N_{1}+1)\,(N_{2}+1)\,(N_{1}+2)\,(N_{2}+2)\,(N_{1}+3)\,(N_{2}+3)}{24\,(N_{\mathrm{s}}+3)\,(N_{\mathrm{s}}+4)\,(N_{\mathrm{s}}+5)\,(N_{\mathrm{s}}+6)}\,Q_{4,3},
\end{align*}
\begin{align*}
B_{-}=-\frac{N_{1}^{2}\,N_{2}^{2}\,(N_{1}-1)\,(N_{2}-1)\,N_{\Delta}^{2}}{(N_{\mathrm{s}}-2)^{2}\,(N_{\mathrm{s}}-1)\,N_{\mathrm{s}}\,(N_{\mathrm{s}}+2)}\,R_{-},
\end{align*}
\begin{align*}
B_{+}=\frac{N_{1}^{2}\,N_{2}^{2}\,(N_{1}+1)\,(N_{2}+1)\,N_{\Delta}^{2}}{(N_{\mathrm{s}}-2)\,N_{\mathrm{s}}\,(N_{\mathrm{s}}+1)\,(N_{\mathrm{s}}+2)^{2}}\,R_{+},
\end{align*}

where

\begin{align*}
    Q_{1,0}=~&N_{1}^{6}-6N_{1}^{5}N_{2}+N_{1}^{4}(3N_{2}^{2}-2)+4N_{1}^{3}N_{2}(5N_{2}^{2}+7)+N_{1}^{2}(3N_{2}^{4}-96N_{2}^{2}+13)\\[1.2ex]
    &~~~+N_{1}(-6N_{2}^{5}+28N_{2}^{3}+2N_{2})+N_{2}^{6}-2N_{2}^{4}+13N_{2}^{2}-12,
\end{align*}
\begin{align*}
    Q_{2,-1}=~&N_{1}^{6}+N_{1}^{5}\,(3-2N_{2})-N_{1}^{4}\,(N_{2}^{2}+17N_{2}+11)+N_{1}^{3}\,(4N_{2}^{3}+14N_{2}^{2}+60N_{2}+17)\\[1.2ex]
    &~~~~~-N_{1}^{2}\,(N_{2}^{4}-14N_{2}^{3}+98N_{2}^{2}+13N_{2}+2)-N_{1}\,(2N_{2}^{5}+17N_{2}^{4}-60N_{2}^{3}+13N_{2}^{2}+4N_{2}+8)\\[1.2ex]
    &~~~~~+(N_{2}-1)\,N_{2}\,(N_{2}^{4}+4N_{2}^{3}-7N_{2}^{2}+10N_{2}+8),
\end{align*}
\begin{align*}
    Q_{2,1}=~&N_{1}^{6}-N_{1}^{5}\,(2N_{2}+3)-N_{1}^{4}\,(N_{2}^{2}-17N_{2}+11)+N_{1}^{3}\,(4N_{2}^{3}-14N_{2}^{2}+60N_{2}-17)\\[1.2ex]
    &~~~~~-N_{1}^{2}\,(N_{2}^{4}+14N_{2}^{3}+98N_{2}^{2}-13N_{2}+2)+N_{1}\,(-2N_{2}^{5}+17N_{2}^{4}+60N_{2}^{3}+13N_{2}^{2}-4N_{2}+8)\\[1.2ex]
    &~~~~~+(N_{2}+1)\,N_{2}\,(N_{2}^{4}-4N_{2}^{3}-7N_{2}^{2}-10N_{2}+8),
\end{align*}
\begin{align*}
    Q_{4,-1}=-\frac{2N_{1}\,N_{2}\,(N_{1}-1)\,(N_{2}-1)}{(N_{\mathrm{s}}-2)^{2}}-\frac{(N_{1}+1)\,(N_{2}+1)\,(N_{1}-2)\,(N_{2}-2)\,(N_{\mathrm{s}}-1)}{(N_{\mathrm{s}}-4)\,(N_{\mathrm{s}}-3)\,(N_{\mathrm{s}}+2)},
\end{align*}
\begin{align*}
    Q_{4,1}=-\frac{2N_{1}\,N_{2}\,(N_{1}+1)\,(N_{2}+1)}{(N_{\mathrm{s}}+2)^{\,2}}-\frac{(N_{1}-1)\,(N_{2}-1)\,(N_{1}+2)\,(N_{2}+2)\,(N_{\mathrm{s}}+1)}{(N_{\mathrm{s}}-2)\,(N_{\mathrm{s}}+3)\,(N_{\mathrm{s}}+4)},
\end{align*}

and

\begin{align*}
    Q_{2,\,0}\,=\,Q_{3,\,-2}\,=\,Q_{3,\,0}\,=\,Q_{3,\,2}\,=\,Q_{4,\,-3}\,=\,Q_{4,\,0}\,=\,Q_{4,\,3}\,=\,R_{-}\,=\,R_{+}\,=\,1,
\end{align*}
along with
\begin{align*}
    A_{3,-1}=A_{3,1}=A_{4,-2}=A_{4,2}=0.
\end{align*}


\bibliography{refs}

\end{document}